\documentclass[aps,twocolumn,prl,superscriptaddress]{revtex4-2}
\usepackage{amssymb}
\usepackage{amsmath}
\usepackage{bbm}
\usepackage{color}
\usepackage{tikz}
\usepackage{appendix}
\usepackage{hyperref}
\hypersetup{
  colorlinks=true,
  citecolor=blue,
  linkcolor=blue,
  urlcolor=blue,
}

\begin{document}
\title{Unified characterization for higher-order topological phase transitions}
\author{Wei Jia}
\affiliation{International Center for Quantum Materials and School of Physics, Peking University, Beijing 100871, China}
\affiliation{Hefei National Laboratory, Hefei 230088, China}
\author{Xin-Chi Zhou}
\affiliation{International Center for Quantum Materials and School of Physics, Peking University, Beijing 100871, China}
\affiliation{Hefei National Laboratory, Hefei 230088, China}
\author{Lin Zhang}
\affiliation{ICFO-Institut de Ciencies Fotoniques, The Barcelona Institute of Science and Technology, Av. Carl Friedrich Gauss 3, 08860 Castelldefels (Barcelona), Spain}
\author{Long Zhang}
\affiliation{School of Physics and Institute for Quantum Science and Engineering, Huazhong University of Science and Technology, Wuhan 430074, China}
\author{Xiong-Jun Liu}
\thanks{Correspondence addressed to: xiongjunliu@pku.edu.cn}
\affiliation{International Center for Quantum Materials and School of Physics, Peking University, Beijing 100871, China}
\affiliation{Hefei National Laboratory, Hefei 230088, China}
\affiliation{International Quantum Academy, Shenzhen 518048, China}

\begin{abstract}
Higher-order topological phase transitions (HOTPTs) are associated with closing either the bulk energy gap (type-I) or boundary energy gap (type-II) without changing symmetry, and conventionally the both transitions are captured in real space and characterized separately. Here we propose a momentum-space topological characterization of the HOTPTs, which unifies the both types of topological transitions and enables a precise detection by quench dynamics. Our unified characterization is based on a novel correspondence between the mass domain walls on real-space boundaries and the higher-order band-inversion surfaces (BIS) which are characteristic interfaces in the momentum subspace. The topological transitions occur when momentum-space topological nodes, dubbed higher-order topological charges, cross the higher-order BISs after proper projection. Particularly, the bulk (boundary) gap closes when all (part of) topological charges cross the BISs, characterizing the type-I (type-II) HOTPTs. These distinct dynamical behaviours of higher-order topological charges can be feasibly measured from quench dynamics driven with control in experiments. Our work opens an avenue to characterize and detect the two types of HOTPTs within a unified framework, and shall advance the research in both theory and experiment.
\end{abstract}

\maketitle

{\em Introduction.}--Higher-order topological phases~\cite{sitte2012topological,zhang2013surface,benalcazar2017quantized,song2017d,wang2018high,yan2018majorana} have drawn widespread attention in recent years. These topologically nontrivial phases generalize the well-known bulk-boundary correspondence, so that a $d$-dimensional ($d$D) $n$th-order topological phase host gapless states in the $(d-n)$D boundary, while its higher-dimensional boundaries are gapped, rendering the key feature of rich new topological states~\cite{slager2015impurity,benalcazar2017electric,liu2017novel,langbehn2017reflection,li2017engineering,schindler2018higher,ezawa2018higher,ezawa2018topological,khalaf2018higher,geier2018second,queiroz2019splitting,sheng2019two,trifunovic2019higher,cualuguaru2019higher,park2019higher,benalcazar2019quantization,yan2019higher,zhu2019second,pan2019lattice,volpez2019second,zeng2019majorana,ren2020engineering,tiwari2020unhinging,zhang2020mobius,lee2020two,trifunovic2021higher,tan2022two}. More recently, the higher-order topological states have also been found in Floquet systems~\cite{peng2019floquet,huang2020floquet,hu2020dynamical}, non-Hermitian systems~\cite{luo2019higher,lee2019hybrid,zhang2019non,liu2019second}, interacting systems~\cite{stepanenko2022higher,may2022interaction}, and fractal systems~\cite{yang2020photonic,zheng2022observation}.

Since the bulk, as well as partially the boundary, is gapped for higher-order topological states, the higher-order topological phase transitions (HOTPTs) are associated with closing either the bulk (type-I) or boundary (type-II) energy gap without changing symmetry~\cite{ghorashi2020vortex,kheirkhah2020first,lichangan2020topological,wu2020boundary,ezawa2020edge,asaga2020boundary,claes2020wannier,khalaf2021boundary,yang2020type,du2022acoustic,mao2022orbital,chen2022experimental}. Currently, the type-I and type-II HOTPTs are characterized by the topological invariants defined on the bulk and Wannier bands~\cite{marzari2012maximally}, respectively. For instance, the multipole moments~\cite{benalcazar2017electric} and bulk polarization~\cite{liu2017novel} can only be applied to identify the type-I HOTPTs, while the nested Wilson loop~\cite{benalcazar2017quantized} and Wannier band polarizations~\cite{khalaf2021boundary} are limited to the type-II HOTPTs. Nevertheless, when the topological transitions occur, these invariants defined by the bulk and boundary properties are not unified and can not fully capture all the transitions~\cite{yang2021hybrid,benalcazar2022chiral,luo2022higher}. Hence the independent characterization can not essentially describe the HOTPTs and is not conducive to uncover the novel higher-order topological states. 

Meanwhile, the current characterizations bring difficulties for identifying the both types of topological transitions in experiments~\cite{cerjan2020observation,niu2020simulation,xie2021higher}. Very recently, the experimental realizations of the higher-order topological states have been widely reported in cleaning synthetic systems in a controllable fashion~\cite{serra2018observation,peterson2018quantized,imhof2018topolectrical,ni2019observation,fan2019elastic,xue2019acoustic,el2019corner,kempkes2019robust,wei20213d}. The bulk physics can be conveniently simulated in synthetic systems like ultracold atoms~\cite{bloch2012quantum,jotzu2014experimental,wu2016realization,schafer2020tools}, nitrogen-vacancy center~\cite{kong2016direct,ariyaratne2018nanoscale,ji2020quantum}, and nuclear magnetic resonance~\cite{xin2020experimental,zhao2021characterizing}, while the classical simulators (such as phononic crystals~\cite{serra2018observation}, photonic crystals~\cite{peterson2018quantized}, and electric circuits~\cite{imhof2018topolectrical}) provide ideal grounds to play with the higher-order boundary modes. However, while having high controllability, it is still challenging for these synthetic systems to observe the two types of HOTPTs due to the lack of a full manipulation and detection of both the bulk and boundary physics. 

Motivated by these considerations, in this Letter, we propose a unified characterization for the both fundamental types of HOTPTs, which goes beyond the traditional independent characterization and enables a feasible detection of HOTPTs via quench dynamics. We first show a generic duality that for a $d$D $n$th-order topological phase, the existence of $(d-n)$D gapless boundary states uniquely corresponds to the emergence of $n$th-order band-inversion surfaces (BISs)~\cite{zhang2018dynamical,song2019observation,ye2020emergent,zhang2020unified,yu2020high,li2020topological,wang2021realization,chen2021digital,zhang2021universal,Lei2022} which are $(d-n)$D interfaces in the momentum space characterizing where the energy bands cross and are inverted. Based on this nontrivial duality, the topological phase transitions occur when the higher-order topological charges cross the higher-order BISs after proper projection, with the type-I (or type-II) transitions being characterized by all (or part of) charges pass through the BISs, providing an elegant and unified characterization of both types of HOTPTs. We finally show that both topological charges and BISs can be well measured in quantum quench experiments. Our work provides a new way to simulate the higher-order topological phases and detect the HOTPTs.

{\em Duality between mass domain wall and BIS.}--We start with deriving a duality between mass domain wall (MDW) and BISs for a generic $d$D $n$th-order topological insulator (TI) captured by the Hamiltonian
\begin{equation}\label{eq1}
\begin{split}
\mathcal{H}_{\mathbf{k}}=\sum^{d}_{j=1}h_j(k_j)\gamma_j+\sum^{n}_{l=1}h_{d+l}(k_1,\cdots,k_{d-l+1})\gamma_{d+l},
\end{split}
\end{equation}
where $\mathbf{k}=(k_1,k_2,\cdots, k_{d})$ is the $d$D momentum.
The Gamma matrices obey the anticommutation relation of Clifford algebra~\cite{morimoto2013topological,chiu2013classification} and can be regarded as the (pseudo)spin operators. Here we use the convention that $h_{j\leqslant d}$ denotes (pseudo)spin-orbit coupling coefficients, while $h_{d+l}$ represents mass terms which include the Zeeman terms. Without mass terms the Hamiltonian $\mathcal{H}_{\mathbf{k}}$ characterizes a massless Dirac semimetal. The mass term for $n=1$ opens a bulk gap and gives the 1st-order topological model, such as the 1D Su-Schrieffer-Heeger (SSH) chain~\cite{su1980soliton} and 2D Haldane model~\cite{haldane1988model}. For $n>1$ the additional mass terms further open gaps on boundary and give rise to the higher-order topological phases, including the 2nd-order TIs with order-two symmetry~\cite{geier2018second} and the 3rd-order TIs with inversion and reflection symmetries~\cite{li2020topological}, where the crystalline symmetries determine the configurations of $(d-n)$D gapless boundary modes.

\begin{figure}[!t]
\includegraphics[width=\columnwidth]{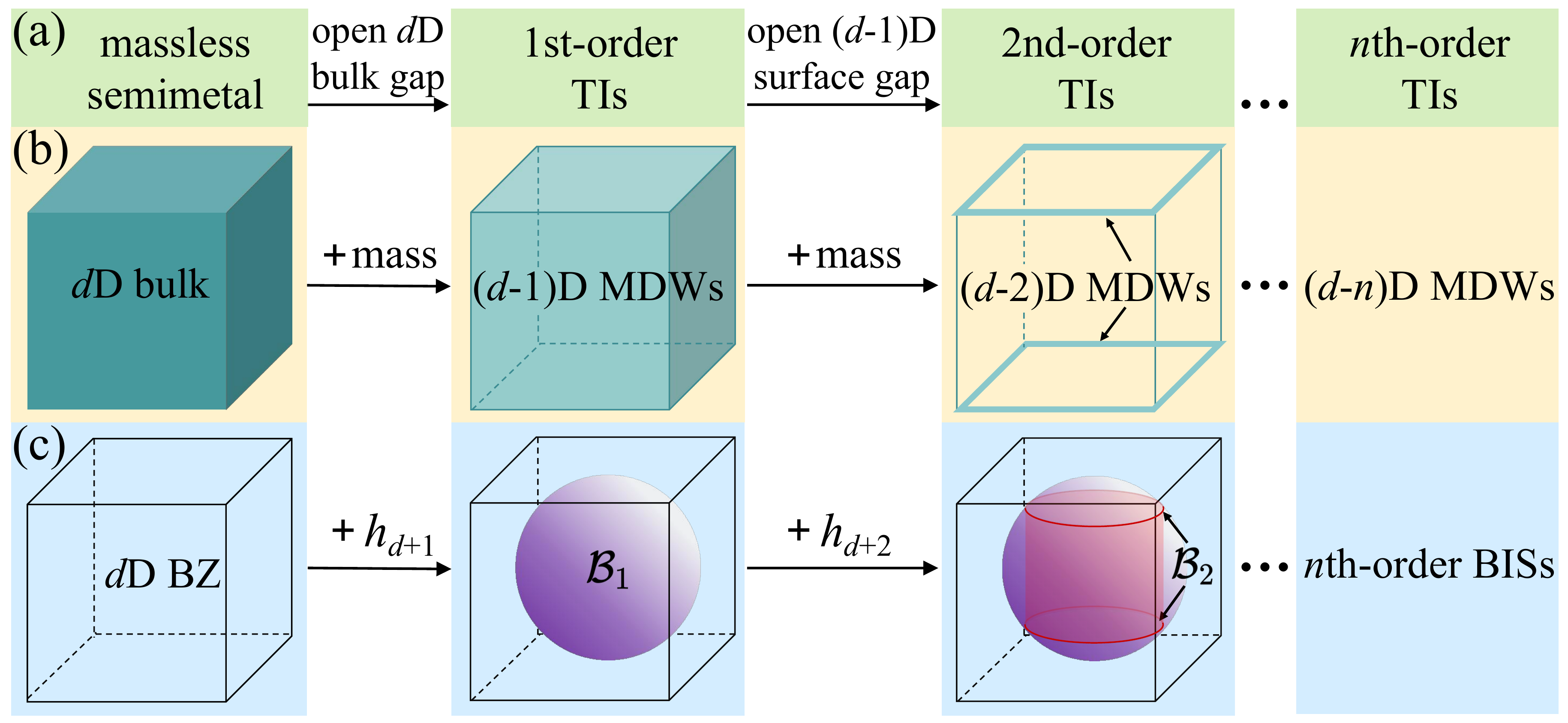}
\caption{Schematic of MDW-BIS duality. (a) Construction of $d$D $n$th-order TIs from massless semimetals by adding additional mass terms. (b) The corresponding $(d-n)$D MDWs in real space. (c) The corresponding $n$th-order BISs $\mathcal{B}_n$ in BZ.}
\label{Fig-1}
\end{figure}

The boundary states of higher-order topological phases can be characterized through the dimensional reduction approach. Namely, the boundary states of an $n$th-order topological phase are obtained as Jackiw-Rebbi modes~\cite{jackiw1976solitons} by introducing Dirac MDWs into the $(d-n+1)$D boundary states of a $(n-1)$th-order topological phase~[see Figs.~\ref{Fig-1}(a) and \ref{Fig-1}(b)]. Accordingly, the $(d-n)$D MDWs in real space can correspond to the momentum-space $(d-n)$D closed surfaces with vanishing mass terms of Hamiltonian (\ref{eq1}), defining the $n$th-order BISs $\mathcal{B}_n\equiv\{\mathbf{k}|h_{d+l}=0,l=1,2,\dots,n\}$~[see Fig.~\ref{Fig-1}(c)]. This renders a MDW-BIS duality for the $n$th-order topological phases. Below we briefly illustrate this duality of the Hamiltonian \eqref{eq1}. More details of the generic proofs are provided in Supplementary Material~\cite{supplementary_material}.

We first start from the 1st-order TIs ($n=1$). The corresponding gapless surface states can be described as bound modes at the $(d-1)$D MDWs between the system and vacuum on real-space boundary. On the other hand, these surface states are uniquely determined by the bulk topology, which is known to be further characterized by the $(d-1)$D 1st-order BIS $\mathcal{B}_1$ in momentum space with vanishing mass term $h_{d+1}(\mathbf{k})=0$~\cite{zhang2018dynamical}. This renders the MDW-BIS duality for the $1$st-order TIs. Further, the 2nd-order topological phase is obtained when an additional mass term $h_{d+2}$ is added to the Hamiltonian of $1$st-order TIs; see Eq.~\eqref{eq1}. The condition $h_{d+2}(\mathbf{k})=0$ gives another 1st-order BIS, and its crossing with the BIS $\mathcal{B}_1$ results in a $(d-2)$D 2nd-order BIS $\mathcal{B}_2$. The existence of $\mathcal{B}_2$ immediately implies that the additional mass term $h_{d+2}$ must have sign changes after projected onto the $(d-1)$D surface states of the $1$st-order TIs~\cite{supplementary_material}. Hence the $h_{d+2}$ term gaps out the $(d-1)$D surface states almost everywhere but leaves the MDWs on the $(d-2)$D boundary, yielding a 2nd-order topological phase. {\color{black}Repeating the above procedures, we obtain all of the higher-order topological phases, rendering the generic duality between the $n$th-order BISs and $(d-n)$D MDWs for the $n$th-order topological phases. This nontrivial duality reveals a new correspondence between the momentum- space bulk physics and the real-space boundary physics. Moreover, it is also faithful for the higher-order topological phases with stacking $1$D SSH chain, such as the $2$D and $3$D BBH models~\cite{benalcazar2017quantized,benalcazar2017electric} which can arrive at Hamiltonian~\eqref{eq1} by rescaling the Gamma matrices~\cite{trifunovic2021higher}.}

{\em Topological characterization of HOTPTs.}--We now develop the unified characterization of HOTPTs based on the above MDW-BIS duality. Here a key idea is that an $n$th-order topological system can be equivalently transformed into the superposition of $n$ effective 1st-order topological subsystems by using the dimensional reduction. Specifically, the $(d-i)$D MDWs are introduced to the $(d-i+1)$D boundary states by adding an additional mass term $h_{d+i}$ to the Hamiltonian with $i=1, 2, \dots, n$, turning the $(i-1)$th-order topological phases into a $i$th-order topological phase. We treat the $(d-i+1)$D gapless boundary modes as a massless Dirac system, and then the MDWs of the $i$th-order topological phases are indeed the boundary states of an effective 1st-order $(d-i+1)$D gapped topological phase given by
\begin{equation}\label{h2}
\mathcal{H}_{\mathbf{k}^{(i-1)}}=\sum_{j\in D^{(i-1)}}h_j(k_j)\gamma^{(i-1)}_j+h_{d+i}(\mathbf{k}^{(i-1)})\gamma^{(i-1)}_{d+i}
\end{equation}
with $D^{(i-1)}$ being a subset of $\{1,2,\dots,d\}$ with $(d-i+1)$ elements~\cite{supplementary_material}. {\color{black}For these effective 1st-order topological subsystems, the $(d-i+1)$D momentum subspace $\mathbf{k}^{(i-1)}$ characterizes an effective $(d-i+1)$D Brillouin zone $\mathrm{BZ}^{(i-1)}$ obtained by projecting $\mathcal{B}_{i-1}$ onto the subspace spanned by all $k_{j}$, with the Gamma matrices $\gamma^{(i-1)}_{j}$ being generally superpositions of the original ones~\cite{notes1}.} 

With above observation, the topological index $\mathcal{V}_{n}$ of the $n$th-order topological phase \eqref{eq1} is then determined by all of the invariants $w_{i}$ of the effective $1$st-order topological Hamiltonian $\mathcal{H}_{\mathbf{k}^{(i-1)}}$, given by
\begin{equation}\label{eq3}
\mathcal{V}_n=\text{sgn}\left(|w_1w_2\cdots w_{n-1}|\right)w_{n}.
\end{equation}
This can be easily understood in process of constructing an $n$th-order topological phase from the $(n-1)$th-order TI with the topological index $\mathcal{V}_{n-1}$. The sign function $\text{sgn}(|\mathcal{V}_{n-1}|)=1$ (or $0$) characterizes the presence (or absence) of the $(d-n+1)$D boundary states for the $(n-1)$th-order topological phase. As indicated by Eq.~\eqref{h2}, the topology of the $n$th-order TI is inherited from these boundary modes~\cite{mong2011edge,kunst2017anatomy} and is characterized by the invariant $w_{n}$, while the absence of these boundary states always leads to a trivial $n$th-order phase. Thus the $n$th-order TI has the topological invariant $\mathcal{V}_{n}=\text{sgn}(|\mathcal{V}_{n-1}|)w_{n}$. Repeating the same analysis for all $\mathcal{V}_{i\leqslant n-1}$ yields the topological index \eqref{eq3}.

The last step for the unified characterization is to represent $w_i$ in terms of topological charges, which are dual to the BISs~\cite{zhang2019characterizing}. For above effective $1$st-order topological Hamiltonian, an $s$th-order topological charge $\mathcal{C}^{(i-1)}_{s,q}=\mathrm{sgn}[J_{\mathbf{h}_{\mathrm{so}}}(\mathbf{k}^{(i-1)}_q)]$ is a nodal point of (pseudo)spin-orbit coupled field $\mathbf{h}_{\mathrm{so}}(\mathbf{k}^{(i-1)})=(h_1,h_2,\cdots,h_{d-i+2-s})$ at momenta $\mathbf{k}^{(i-1)}_q$ and quantified by Jacobian determinant $J_{\mathbf{h}_{\mathrm{so}}}(\mathbf{k}^{(i-1)})\equiv\mathrm{det}(\partial h_{\mathrm{so},{j'}}/\partial k_{j'})$~\cite{jia2020charge,supplementary_material}. Accordingly, the invariant $w_i$ equals the total monopole charges enclosed by $\mathcal{B}^{(i-1)}_{\mathrm{proj},s}\equiv\{\mathbf{k}^{(i-1)}|h_{d-i+3-s}=\cdots=h_{d+i}=0\}$, dubbed as the \emph{projective} $s$th-order BISs. We then obtain
\begin{equation}\label{eq:wi}
w_{i}=\sum_{q\in\bar{\mathcal{B}}^{(i-1)}_{\text{proj},s}}\mathcal{C}^{(i-1)}_{s,q},
\end{equation}
where $\bar{\mathcal{B}}^{(i-1)}_{\mathrm{proj},s}$ is the momentum region enclosed by $\mathcal{B}^{(i-1)}_{\mathrm{proj},s}$ with $h_{d-i+3-s}<0$. 
Unlike the higher-order BISs in the original bulk system, these projective higher-order BISs are defined in the effective $\mathrm{BZ}^{(i-1)}$, since $\mathcal{B}^{(i-1)}_{\mathrm{proj},s}$ is actually the projection of the $i$th-order BIS $\mathcal{B}_{i}$ onto $\text{BZ}^{(i-1)}$. The Eqs.~\eqref{eq3} and~\eqref{eq:wi} give the characterization for a broad class of higher-order topological phases with various lattice symmetries and described by Hamiltonian~\eqref{eq1}. We show later that while the characterization is built on the topological indices $w_i$ of the effective $1$st-order topological system, it can be precisely measured in experiment by quench dynamics. 

\begin{figure}[!tbp]
\includegraphics[width=\columnwidth]{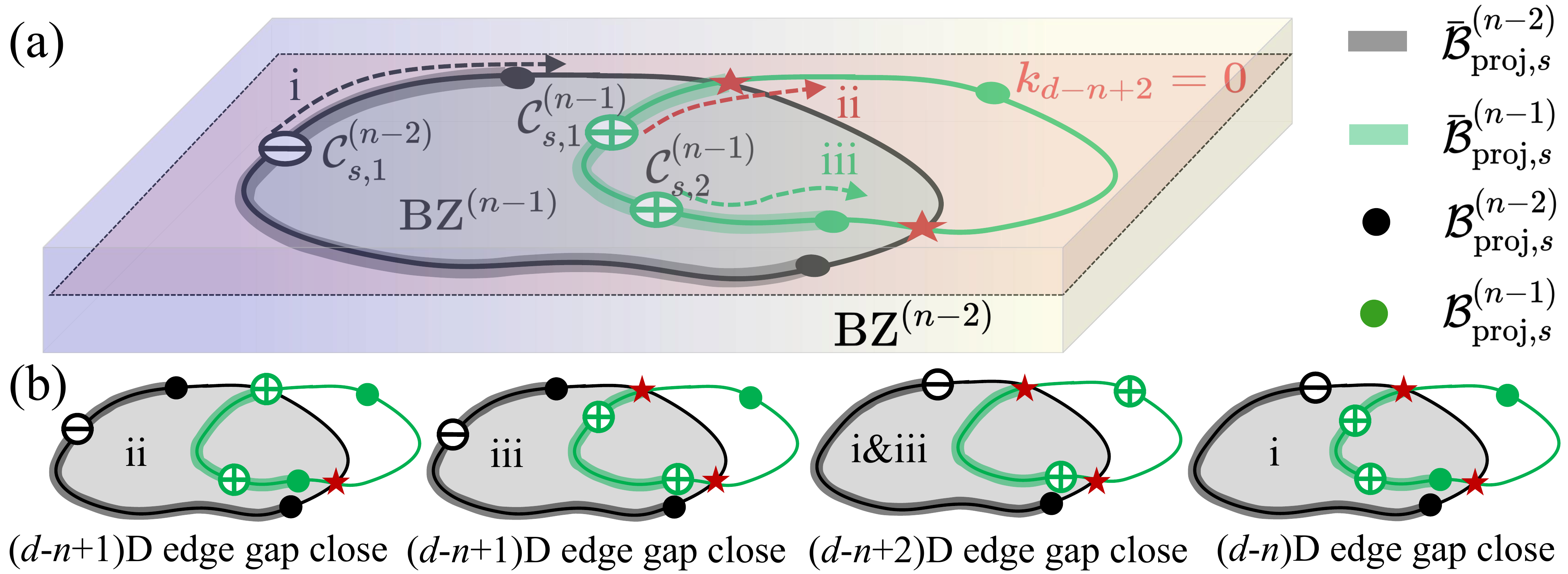}
\caption{Schematic of HOTPTs. (a) Behaviour of topological charges in the phase transitions. Here the charges cross either the projective BISs [black (i) and green (iii) dots] or the border [red stars (ii)] of $\text{BZ}^{(n-1)}$ (gray regions). The boundary gaps for $i<n-1$ are assumed to be open. (b) In ii or iii, driving an $(n-1)$th-order topological phase. While i together with iii gives an $(n-2)$th-order topological phase. The $n$th-order topological phase shall remains unchanged when only i occurs.
}
\label{Fig-2}
\end{figure}

\begin{figure*}[!htbp]
\includegraphics[width=\textwidth]{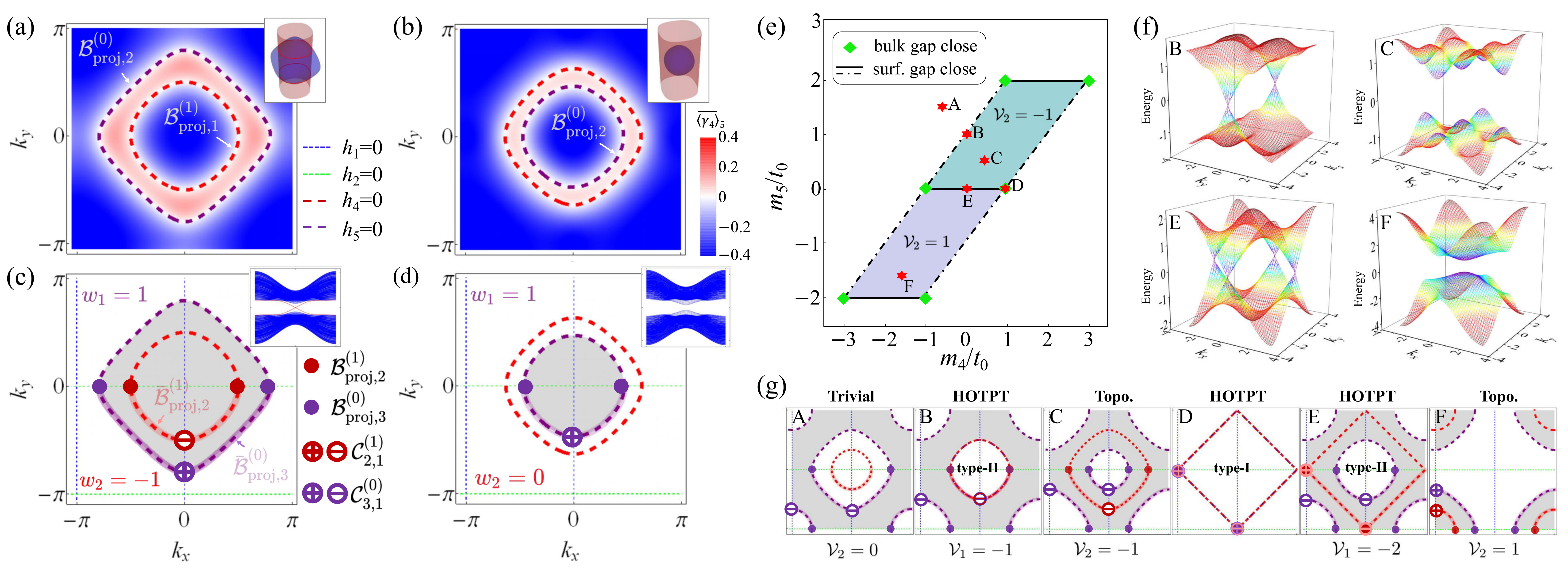}
\caption{Numerical results of 3D 2nd-order TI. (a)-(b) Time-averaged spin texture of $\overline{\langle\gamma_4(\mathbf{k}) \rangle}_{5}$ at $k_z=0$ by quenching each $h$-axis from $\delta m_{1,2,3,4,5}=30t_0$ to $0$, where $(m_4,m_5)=(1.2t_0,t_0)$ is for (a) and $(2.1t_0,0.6t_0)$ for (b). The vanishing polarization gives $\mathcal{B}^{(1)}_{\text{proj},1}$ and $\mathcal{B}^{(0)}_{\text{proj},2}$ in (a), while there is no $\mathcal{B}^{(1)}_{\text{proj},1}$ in (b) since the red dashed-ring is not in $\text{BZ}^{(1)}$ (gray regions). The insets give $\mathcal{B}_1$ (a blue spherical) and $\mathcal{B}_2$ (two red rings) in original momentum space. (c-d) Configurations of the charges and BISs for (a) and (b), where the insets are $k_yk_z$-OBC spectra. (e) Phase diagram with the parameter points $A-F$ (red stars). (f) The two lowest $2$D surface energies. (g) Configurations of the charges and BISs for $A-F$.
}
\label{Fig-3}
\end{figure*}

We are now ready to write down the unified characterization of the HOTPTs which must be associated with the change of $w_{i}$ for one or multiple effective $1$st-order topological subsystems~\eqref{h2}. Equivalently, in a HOTPT the topological charges $\mathcal{C}^{(i-1)}_{s,q}$ must cross either the projective BIS $\mathcal{B}^{(i-1)}_{\text{proj},s}$ or the border of $\text{BZ}^{(i-1)}$ [see Fig.~\ref{Fig-2}(a)]. Namely, a topological transition of $n$th-order phase
\begin{equation}\label{tpt}
\mathcal{V}_n\xrightarrow[\text{type-I}(m=1)]{\text{type-II}(m>1)}\mathcal{V}_{m-1}\longrightarrow\mathcal{V}_{p}
\end{equation}
occurs for $i$ taking values from $m$ to $n$, yielding at the critical point $(d-m+1)$D gapless boundaries characterized by $\mathcal{V}_{m-1}$ [see Fig.~\ref{Fig-2}(b)]. For $m=1$, all topological charges cross the projective BISs, and the bulk gap closes at the critical point, manifesting the type-I transition. For $m>1$, only part of the topological charges cross the projective BISs (or the border of effective BZ). Accordingly, the energy gap closes only for the $(d-m+1)$D boundaries parallel (or perpendicular) to the lower-dimensional ${\rm BZ}^{(m-1)}$, giving a type-II transition. This characterization also further precisely classifies the type-II transition into different $m$-orders~\cite{supplementary_material}. Moreover, it characterizes the HOTPT between an initial $n$-order phase and a final $p$th-order phase with $p\leqslant n$, given that $w_{i\leqslant p}$ is nonzero according to Eq.~\eqref{eq3}.

{\em Dynamical detection and Application.}--We show now that the unified characterization can facilitate the precise detection of the HOTPTs and propose the applications based on quantum quenches. Our scheme is based on sequentially quenching all of the (pseudo)spin axes $\gamma_{\alpha=1,2,\dots,d+n}$, while only measuring a single (pseudo)spin component $\gamma_{d+1}$ in each quench. For this we suddenly tune the Hamiltonian $\mathcal{H}_{\mathbf{k}}+\delta m_{\alpha}\gamma_{\alpha}$ from deep trivial regime $|\delta m_{\alpha}|\gg |h_{\alpha}|$ to topological regime $\delta m_{\alpha}=0$. The time-averaged (pseudo)spin polarization $\gamma_{d+1}$ after quench is given by $\overline{\langle\gamma_{d+1}(\mathbf{k}) \rangle}_{\alpha}\equiv\lim_{T\to\infty}{1\over T}\int^{T}_0\text{Tr}[\rho_{\alpha} e^{\mathrm{i}\mathcal{H}_{\mathbf{k}}t} \gamma_\alpha e^{-\mathrm{i}\mathcal{H}_{\mathbf{k}}t}]\text{d}t$, where $\rho_{\alpha}$ is density matrix for initial state. The projective BISs are determined as
$\mathcal{B}^{(i-1)}_{\text{proj},s}=\{\mathbf{k}^{(i-1)}|\overline{\langle\gamma_{d+1}\rangle}_{d-i+3-s}=\cdots=\overline{\langle\gamma_{d+1}\rangle}_{d+i}=0\}$. The higher-order topological charge $\mathcal{C}^{(i-1)}_{s,q}$ is further detected by the dynamical field
\begin{equation}\label{de}
\begin{split}
g_{j'}=-\lim_{\mathbf{k}^{(i-1)}\rightarrow \mathbf{k}^{(i-1)}_q}\frac{\text{sgn}\left(h_{\beta}\right)}{{\mathcal{N}_{\mathbf{k}^{(i-1)}}}}\frac{\overline{\langle\gamma_{d+1}\rangle}_{j'}\overline{\langle\gamma_{d+1}\rangle}_{\beta}}{\overline{\langle\gamma_{d+1}\rangle}_{d+i}}
\end{split}
\end{equation}
for $s=1$ with $\beta=d+i$ and $s>1$ with $\beta=d-i+2-s$, since it can be shown that $g_{j'}=h_{\mathrm{so},j'}$ near the node point $\mathbf{k}^{(i-1)}_{q}$. Here $\mathcal{N}_{\mathbf{k}^{(i-1)}}$ is a normalization factor. This dynamical detection scheme is highly feasible in experiment as we demonstrate below.

We exemplify the application of the unified characterization with a $3$D $2$nd-order TI, constructed by adding a mass term into the 3D chiral TI~\cite{ji2020quantum,xin2020experimental}, with the bulk Hamiltonian $\mathcal{H}_{\mathbf{k}}=\sum^{5}_{\alpha =1}h_\alpha\gamma_\alpha=h_1\sigma_x\tau_0+h_2\sigma_y\tau_0+h_3\sigma_z\tau_x+h_4\sigma_z\tau_z+h_5\sigma_z\tau_y$, where $h_{1,2,3}=t_{\text{so}}\sin k_{x,y,z}$, $h_4=m_4-t_0(\cos k_x+\cos k_y+\cos k_z)$, and $h_5=m_5-t_0(\cos k_x+\cos k_y)$. Here $\boldsymbol{\sigma}$ and $\boldsymbol{\tau}$ are both Pauli matrices and $k_{1,2,3}=k_{x,y,z}$. From the time-averaged spin texture shown in Fig.~\ref{Fig-3}(a), we observe a ring-shaped projective BIS $\mathcal{B}^{(1)}_{\text{proj},1}$, manifesting the existence of the BIS $\mathcal{B}_2$ in original momentum space and identifying the emergence of hinge states according to the MDW-BIS duality. Moreover, one negative (positive) topological charge $\mathcal{C}^{(1)}_{2,1}$ [$\mathcal{C}^{(0)}_{3,1}$] is observed in the region $\bar{\mathcal{B}}^{(1)}_{\text{proj},2}$ [$\bar{\mathcal{B}}^{(0)}_{\text{proj},3}$] [see Fig.~\ref{Fig-3}(c)], giving the topological invariant $\mathcal{V}_2=-1$. Then the two-fold degenerate zero energy states are localized at the hinges of the top and bottom surfaces along $z$-direction, and protected by the $C^{z}_4$-rotation symmetry
and the anti-reflection symmetry $\bar{\mathcal{R}}_j^{\dagger}\mathcal{H}_{k_j}\bar{\mathcal{R}}_j=-\mathcal{H}_{-k_j}$ along the $j=x,y,z$ axis. However, when $\mathcal{B}_2$ disappears [see Fig.~\ref{Fig-3}(b)], there is no $\mathcal{B}^{(1)}_{\text{proj},1}$ and topological charge $\mathcal{C}^{(1)}_{2,q}$ in $\text{BZ}^{(1)}$ [see Fig.~\ref{Fig-3}(d)]. Hence we have $w_2=0$ and no zero energy state exists in the hinges.

Based on the MDW-BIS duality, the existence of $\mathcal{B}_2$ gives the 2nd-order topological phase diagram $0<|m_5|<2t_0$ and $|m_4-m_5|<t_0$, as shown in Fig.~\ref{Fig-3}(e). One can see that the bulk energy bands become gapless at $(m_4,m_5)=\pm(3t_0,2t_0), \pm(t_0,2t_0), \pm(t_0,0)$ (green squares), while the surface energy gap is only closed at $m_5=0,\pm 2t_0$ (solid lines) and $m_4=m_5\pm t_0$ (dot-dashed lines), which are confined to the real-space interfaces along and perpendicular to $z$ direction, respectively~\cite{supplementary_material}. We shall choose two different parameter paths to observe the phase transitions. In path $A$-$B$-$C$-$D$ [see Fig.~\ref{Fig-3}(e)], there is one negative topological charge crossing the border (purple dashed curves) of $\text{BZ}^{(1)}$ [see Fig.~\ref{Fig-3}(g)], and the surface energy gap closes in both $xz$ and $yz$ planes for $B$ [see Fig.~\ref{Fig-3}(f)]. This renders a type-II transition with $\mathcal{V}_2=0\rightarrow\mathcal{V}_1=-1\rightarrow\mathcal{V}_2=-1$ from $A$ to $C$. As $m_4$ is further increased, all topological charges simultaneously move to the projective BISs for $D$, then the bulk energy gap closes, rendering a type-I transition. In another path $C$-$E$-$F$ [see Fig.~\ref{Fig-3}(e)], the surface energy gap closes in the $xy$ plane for $E$. There is one negative topological charge crossing the projective BISs and changed into a positive topological charge. The higher-order topological transition occurs as $\mathcal{V}_2=-1\rightarrow\mathcal{V}_1=-2\rightarrow\mathcal{V}_2=1$. The unified characterization has explicit advantages that the topological phase transitions of different types can be resolved in quench dynamics.

{\em Discussion and Conclusion.}--The unified characterization also shows that the type-II transitions are further classified into different $m$-orders which can be precisely determined by quench detection. In Supplementary Material~\cite{supplementary_material}, we have presented more relevant examples for the 2D 2nd-order and 3D 3rd-order TIs, which further showcase the broad applicability of the unified characterization. Moreover, our unified theory may be applied to study the Floquet higher-order phases and phase transitions, such as clarifying which type of phase transitions dominates the emergence of Floquet corner modes~\cite{zhou2022generating}. This shall further promote the study of topological phase transitions in Floquet higher-order systems.

In summary, we have shown a unified characterization in momentum space for the higher-order topological phase transitions and further proposed the detection by quench dynamics. The unified characterization is built on the MDW-BIS duality which relates the higher-order boundary modes in real space and the higher-order BISs with topological charges in the momentum space. The topological phase transitions of two types and various orders are generically identified by the higher-order topological charges crossing over the BISs after proper projection, which can be precisely detected by quench dynamics. This work establishes a unified and fundamental characterization of the higher-order topological phases and phase transitions, and shall advance the further broad studies in theory and experiment.

\par This work was supported by National Key Research and Development Program of China (2021YFA1400900), the National Natural Science Foundation of China (Grants No. 11825401, No. 12261160368, and No. 11921005), and the Innovation Program for Quantum Science and Technology (Grant No. 2021ZD0302000). Long Zhang also acknowledges support from the startup grant of the Huazhong University of Science and Technology (Grant No. 3004012191). Lin Zhang also acknowledges support from: ERC AdG NOQIA; Ministerio de Ciencia y Innovation Agencia Estatal de Investigaciones (PGC2018-097027-B-I00/10.13039/501100011033, CEX2019-000910-S/10.13039/501100011033, Plan National FIDEUA PID2019-106901GB-I00, FPI, QUANTERA MAQS PCI2019-111828-2, QUANTERA DYNAMITE PCI2022-132919, Proyectos de I+D+I ``Retos Colaboraci{\' o}n'' QUSPIN RTC2019-007196-7); MICIIN with funding from European Union NextGenerationEU(PRTR-C17.I1) and by Generalitat de Catalunya; Fundaci{\' o} Cellex; Fundaci{\' o} Mir-Puig; Generalitat de Catalunya (European Social Fund FEDER and CERCA program, AGAUR Grant No. 2021 SGR 01452, QuantumCAT{\textbackslash}U16-011424, co-funded by ERDF Operational Program of Catalonia 2014-2020); Barcelona Supercomputing Center MareNostrum (FI-2022-1-0042); EU Horizon 2020 FET-OPEN OPTOlogic (Grant No. 899794); EU Horizon Europe Program (Grant Agreement 101080086 --- NeQST), National Science Centre, Poland (Symfonia Grant No. 2016/20/W/ST4/00314); ICFO Internal ``QuantumGaudi'' project; European Union's Horizon 2020 research and innovation program under the Marie-Sk{\l}odowska-Curie grant agreement No. 101029393 (STREDCH) and No. 847648 (``La Caixa'' Junior Leaders fellowships ID100010434: LCF/BQ/PI19/11690013, LCF/BQ/PI20/11760031, LCF/BQ/PR20/11770012, LCF/BQ/PR21/11840013). Views and opinions expressed in this work are, however, those of the author(s) only and do not necessarily reflect those of the European Union, European Climate, Infrastructure and Environment Executive Agency (CINEA), nor any other granting authority. Neither the European Union nor any granting authority can be held responsible for them.

\bibliographystyle{apsrev4-1}

\pagebreak
\clearpage
\onecolumngrid
\flushbottom
\begin{center}
\textbf{\large Supplementary Material for ``Unified characterization for higher-order topological phase transitions"}
\end{center}
\setcounter{equation}{0}
\setcounter{figure}{0}
\setcounter{table}{0}
\makeatletter
\renewcommand{\theequation}{S\arabic{equation}}
\renewcommand{\thefigure}{S\arabic{figure}}
\renewcommand{\bibnumfmt}[1]{[S#1]}
\renewcommand{\citenumfont}[1]{S#1}

In this Supplementary Material, we provide the detailed proof for the duality between mass domain wall (MDW) and band inversion surface (BIS), i.e., the MDW-BIS duality. We also provide the details of characterization for higher-order topological phase transitions (HOTPTs) and the completely numerical results for the dynamical characterization of 2D 2nd-order, 3D 2nd-order, and 3D 3rd-order topological phases.

\subsection{I. MDW-BIS duality and its applications}
\subsubsection{1. $1$st-order topological phases}
\par Firstly, we demonstrate the MDW-BIS duality of 1st-order topological phases. Our starting point is a 1st-order topological insulator (TI) or a 1st-order topological superconductivity (TSC) obeying the lattice Hamiltonian~\cite{chiu2013classification-s}
\begin{equation}\label{eqs1}
\mathcal{H}_{d\text{D}}(\mathbf{k})=\sum^{d}_{j=1}h_j\gamma_j+h_{d+1}\gamma_{d+1},
\end{equation}
where $\mathbf{k}=(k_1,k_2,\cdots, k_d)$ is $d$-dimensional ($d$D) momentum and Gamma matrices $\boldsymbol{\gamma}$ satisfy the anticommutation relation of Clifford algebra. Here $d$ is the spatial dimension of the system and $k_j$ is the momentum in the $j$-th direction. For convenience, we use the convention that $h_j(k_j)$ are (pseudo)spin-orbit (SO) coupling components while $h_{d+1}(k_1,k_2,\cdots,k_{d})$ denotes the mass term with Zeeman coupling constant.

To capture the MDWs of this 1st-order topological system, we take $\mathbf{k}\rightarrow \mathbf{0}$ for the Hamiltonian~\eqref{eqs1} and obtain its low-energy effective model $H_{d\text{D}}(\mathbf{k})=M_{d+1}\gamma_{d+1}+\sum^{d}_{j=1}k_j\gamma_j$ with $M_{d+1}=\tilde{m}_{d+1}-\sum^{d}_{j=1}k^2_j$, where $h_j(k_j\rightarrow 0)=k_j$ and $h_{d+1}(\mathbf{k}\rightarrow \mathbf{0})=M_{d+1}$. Now one can consider a MDW in the $r_d$ direction, i.e., $\tilde{m}_{d+1}=m_0$ ($-m_0$) is a positive (negative) constant effective mass in the region $r_d>0$ ($r_d<0$), which is a topological (trivial) phase. Thus $k_d$ is not a good quantum number and can be replaced by $-\mathrm{i}\partial / \partial_{r_d}$. The low-energy effective model can be rewritten as
\begin{equation}\label{eqs2}
H_{d\text{D}}(k_1,\cdots,k_{d-1},-\mathrm{i}\partial_{r_d})=\gamma_{d+1}\left(M_{d+1}-\mathrm{i}\gamma_{d+1}\gamma_d\frac{\partial}{\partial_{r_d}}\right)+\sum^{d-1}_{j=1}k_j\gamma_j,
\end{equation}
with $M_{d+1}=\tilde{m}_{d+1}+{\partial^2}/{\partial^2_{r_d}}-\sum^{d-1}_{j=1}k^2_j$.
By further taking $\mathrm{i}\gamma_{d+1}\gamma_d\phi=-\phi$ and solving
$\left(M_{d+1}-\mathrm{i}\gamma_{d+1}\gamma_d{\partial}/{\partial_{r_d}}\right)\Phi=0$, the normalized zero-energy modes are obtained as
\begin{equation}\label{eqs3}
\begin{split}
\Phi{(r_d\geqslant 0)}=c_1\left[e^{-\frac{1}{2}(1+M_-)r_d}-e^{-\frac{1}{2}(1-M_-)r_d}\right]\phi ,~~
\Phi{(r_d<0)}=c_2e^{-\frac{1}{2}(1-M_+)r_d}\phi ,\\
\end{split}
\end{equation}
with $M_{\pm}=\sqrt{1\pm 4m_0+4\sum^{d-1}_{j=1}k^2_j}$. Here the boundary conditions are $\Phi(r_d=0)=\Phi(r_d=\pm\infty)=0$, where the normalized parameters are $c_{1}={\sqrt{1-M_-^2}}/(\sqrt{2}M_-)$ and $c_2=\sqrt{M_+-1}$. Since our focus is on the low-energy spectrum, we have the relation $m_0>\sum^{d-1}_{j=1}k^2_j$. Moreover, both $\Re (M_-)<1$ and $M_+>1$ have ensured that the wave functions are normalized, i.e., $\int^{+\infty}_{0}|\Phi{(r_d\geqslant 0)}|^{2}\text{d}r_d=1$ and $\int^{0}_{-\infty}|\Phi{(r_d<0)}|^2\text{d}r_d=1$. Based on these zero-energy modes in the Eq.~\eqref{eqs3}, we observe that this MDW induces the gapless edge states which are located at the $(d-1)$D interfaces perpendicular to $r_d$ direction.

On the other hand, since $\mathrm{i}\gamma_{d+1}\gamma_{d}$ commutes with $\gamma_{j\neq d+1,d}$, these $(d-1)$D edge states can also be characterized by a projective Hamiltonian~\cite{chiu2013classification-s}. Namely, we use the projection operator $\mathbf{P}^{(0)}=(1-\mathrm{i}\gamma_{d+1}\gamma_d)/2$ in the eigenspace of $\mathrm{i}\gamma_{d+1}\gamma_d=-1$ and project $H_{dD}(\mathbf{k})$ into $(d-1)$D edge Hamiltonian
\begin{equation}\label{eqs4}
H^{\text{edge}}_{(d-1)\text{D}}(k_1,\cdots,k_{d-1})=\mathbf{P}^{(0)^{\dagger}}H_{d\text{D}}(\mathbf{k})\mathbf{P}^{(0)}=\sum^{d-1}_{j=1}k_j\gamma^{(1)}_j,
\end{equation}
where $\gamma^{(1)}_j=\mathbf{P}^{(0)^{\dagger}}\gamma_j\mathbf{P}^{(0)}$ is denoted as the Gamma matrices in $(d-1)$D subspace and has the half size of $\gamma_j$. Obviously, the $(d-1)$D edge energy spectrum is $\mathcal{E}^{\text{edge}}_{(d-1)\text{D},\pm}=\pm\sqrt{\sum^{d-1}_{j=1}k^2_j}$ and shows the gapless behaviour. Since the Eq.~\eqref{eqs4} only takes open boundary condition along $k_d$ direction ($k_d$-OBC), the $(d-1)$D edge states are located at the $(d-1)$D interfaces which are perpendicular to $r_d$ direction. When taking the OBCs along each $k_{1},\cdots,{k_d}$ and projecting $H_{dD}(\mathbf{k})$, there are $d$ edge Hamiltonians are similar to Eq.~\eqref{eqs4}, showing that all $(d-1)$D real-space interfaces host the zero-energy edge states. Correspondingly, the effective mass vanishes in the all $(d-1)$D interfaces, which renders $\tilde{m}_{d+1}=0$ and $M_{d+1}=0$ due to $\mathbf{k}\rightarrow \mathbf{0}$. The physical consequence is that $(d-1)$D MDWs are emerged on all boundaries, which is similar to the interfaces between the material and vacuum. 

\begin{figure}[!tbp]
\begin{center}
\includegraphics[width=\columnwidth]{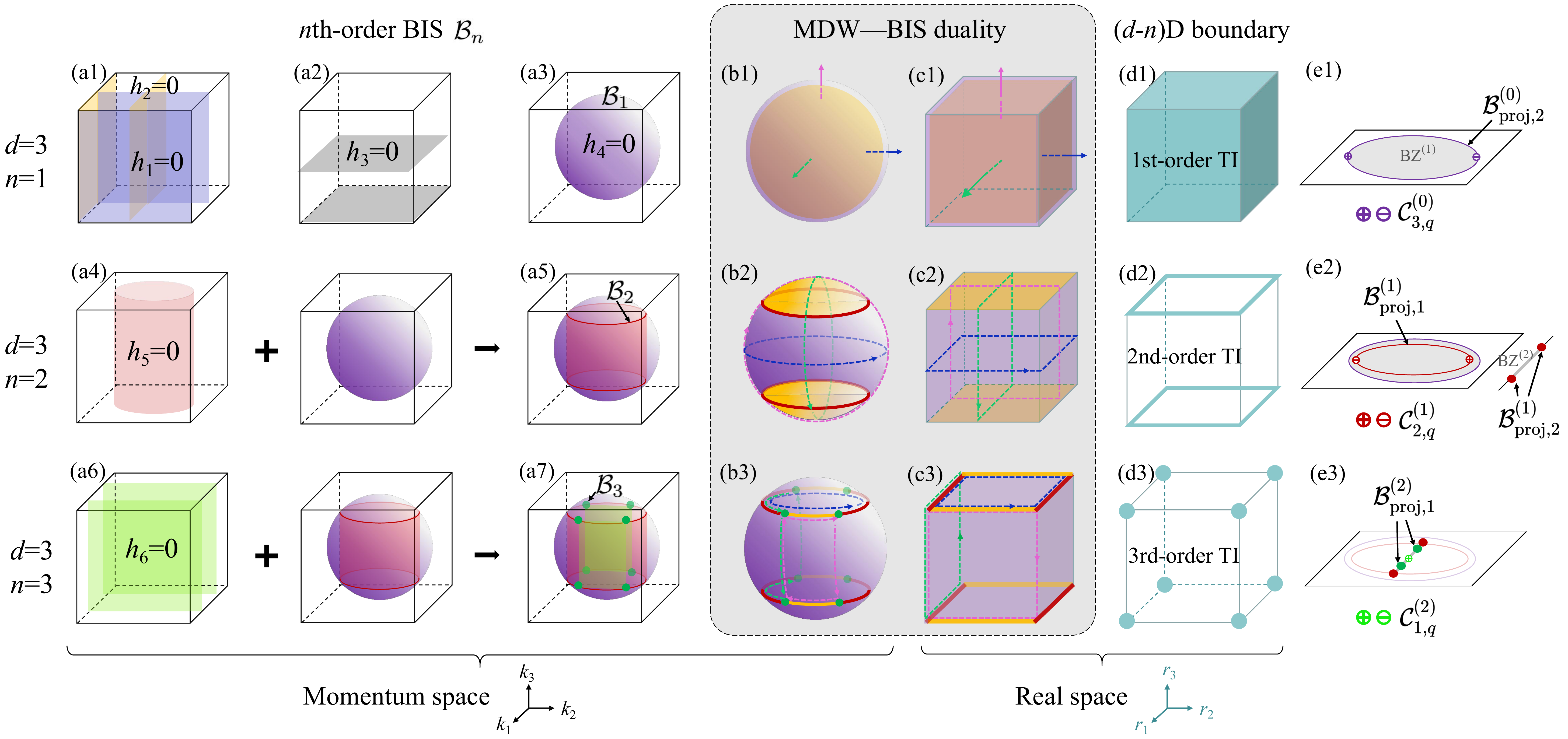}
\caption{MDW-BIS duality for 1st, 2nd, and 3rd-order topological phases in a $3$D system. (a1)-(a7) Sketches of the emergence of 1st, 2nd, and 3rd-order BISs in original momentum space, being denoted as a sphere $(h_4=0)$, two ring-shaped lines $(h_4=h_5=0)$, and eight points $(h_4=h_5=h_6=0)$ respectively. The zero-value SO coupling components $h_1=0$, $h_2=0$, and $h_3=0$ are 2D planes of the original BZ.
(b)-(c) Sketches of the duality between $2$D surface states, $1$D hinge states, and $0$D corner states in real space and $1$st-order, $2$nd-order, and $3$rd-order BISs in momentum space, where the orange and purple regions denotes the different sign for the effective mass in real space and for the mass term in outside and inside of $\mathcal{B}_n$ in momentum space, respectively. The three paths denoted by green, blue, and pink dashed line give that the signs of effective mass of real space and mass term of momentum space have the same changing.
(d1)-(d3) Sketches of $2$D surface states, $1$D hinge states, and $0$D corner states in real space, being denoted as the blue surfaces, blue thick lines, and blue points respectively.
(e1)-(e3) The positive and negative topological charges ``$\oplus$" and ``$\ominus$" live in the effective Brillouin zones $\text{BZ}^{(i-1)}$ (gray regions)
}
\label{Figs1}
\end{center}
\end{figure}

Now we return to the lattice Hamiltonian $\mathcal{H}_{d\text{D}}(\mathbf{k})$. After using the projection operator $\mathbf{P}^{(0)}=(1-\mathrm{i}\gamma_{d+1}\gamma_d)/2$ in the eigenspace of $\mathrm{i}\gamma_{d+1}\gamma_d=-1$, we obtain a $(d-1)$D projective Hamiltonian $\mathbf{P}^{(0)^{\dagger}}\mathcal{H}_{d\text{D}}(\mathbf{k})\mathbf{P}^{(0)}$, which is explicitly written as
\begin{equation}\label{eqs4-sh}
\mathcal{H}^{\text{proj}}_{(d-1)\text{D}}(k_1,\cdots,k_{d-1})=\sum^{d-1}_{j=1}h_j\gamma^{(1)}_j.
\end{equation}
This projective Hamiltonian essentially describes a gapless subsystem in the momentum subspace perpendicular to $k_d$. Since its low-energy effective model is exactly equivalent to Eq.~(\ref{eqs4}), this projective Hamiltonian $\mathcal{H}^{\text{proj}}_{(d-1)\text{D}}$ is topologically equivalent to $H^{\text{edge}}_{(d-1)\text{D}}$ and can characterize the $(d-1)$D edge states perpendicular to $r_d$ direction. Besides, the other $(d-1)$D projective Hamiltonians along $k_{d'\neq d}$-OBC with $d'=1,2,\cdots,d-1$ have similar forms with Eq.~(\ref{eqs4-sh}), and they characterize the $(d-1)$D edge states perpendicular to $r_{d'}$ direction. With this observation, only the mass term $h_{d+1}(\mathbf{k} \rightarrow \mathbf{0})$ gives all $M_{d+1}$-terms in the low-energy model, thus these MDWs with $M_{d+1}=0$ on all $(d-1)$D real-space interfaces correspond to the $(d-1)$D momentum-space surfaces where $h_{d+1}(\mathbf{k})$-term vanishes, defined as the 1st-order BISs $\mathcal{B}_1\equiv \{\mathbf{k}\in \text{BZ}|h_{d+1}(\mathbf{k})=0\}$. This gives the MDW-BIS duality for 1st-order topological phases, showing that the existence of $(d-1)$D gapless boundary states uniquely corresponds to the emergence of $1$st-order BIS [see Figs.~\ref{Figs1}(a3), \ref{Figs1}(b1), \ref{Figs1}(c1), and \ref{Figs1}(d1)]. Moreover, the numbers of $(d-1)$D edge states are characterized by a topological invariant $w_1$, defining as a winding on the 1st-order BIS $\mathcal{B}_1$~\cite{zhang2018dynamical-s}.

From above MDW-BIS duality of the $1$st-order TIs, we also see that the two $h$-components $h_{d}(k_d)$ and $h_{d+1}(\mathbf{k})$ vanish in the projection process of $k_d$-OBC. This implies that the $(d-1)$D edge states, perpendicular to the $r_d$ direction, are only characterized by the projection of 1st-order BISs $\mathcal{B}_1$ on the surfaces of $h_{d}(k_d)=0$, defining an effective Brillouin zone $\text{BZ}^{(1)}$ with the momenta $\mathbf{k}^{(1)}=(k_{1},\cdots,k_{d-1})$. Correspondingly, $\text{BZ}^{(1)}$ shall be limited within the projective 2nd-order BISs $\mathcal{B}^{(0)}_{\text{proj},2}\equiv \{\mathbf{k}|h_{d+1}=h_{d}=0\}$ [see Fig.~\ref{Figs1}(e1)]. It should be emphasized that the so-called projective $s$th-order BIS $\mathcal{B}^{(i-1)}_{\text{proj},s}$ is actually the $s$th-order extension of $\mathcal{B}^{(i-1)}_{\text{proj},1}$. Here $\mathcal{B}^{(i-1)}_{\text{proj},1}$ is the projection of $\mathcal{B}_i$ in $\mathbf{k}^{(i-1)}$ with $i=1,2,\cdots, n$. For example of the 1st-order topological phase with $n=1$, $\mathcal{B}^{(0)}_{\text{proj},1}=\mathcal{B}_1$ is located at the original BZ with the momenta $\mathbf{k}^{(0)}=\mathbf{k}$ [i.e. $\text{BZ}^{(0)}$]. The projective $s$th-order BISs $\mathcal{B}^{(0)}_{\text{proj},s}\equiv \{\mathbf{k}|h_{d+1}=h_{d}=\cdots=0\}$ is defined by using $s$ terms of $h$-components of $\mathcal{H}_{d\text{D}}(\mathbf{k})$, which has no essential difference with the definition of higher-order BISs in Ref.~\cite{yu2020high-s}. Hence $w_1$ can also be further treated as an integer invariant on $\mathcal{B}^{(0)}_{\text{proj},s}$~\cite{yu2020high-s}.

\subsubsection{2. $2$nd-order topological phases}

We add a mass term $h_{d+2}(k_1,\cdots,k_{d-1})$ in Hamiltonian (\ref{eqs1}) and obtain
\begin{equation}\label{eqs4-2}
\mathcal{H}_{d\text{D}}(\mathbf{k})=\sum^{d}_{j=1}h_j\gamma_j+h_{d+1}\gamma_{d+1}+h_{d+2}\gamma_{d+2},
\end{equation}
where the Gamma matrices still satisfy the Clifford algebra. This Hamiltonian~\eqref{eqs4-2} describes the $d$D $2$nd-order topological phases. To capture the MDWs of this $2$nd-order topological system~\eqref{eqs4-2}, we still consider its low-energy effective model $H_{d\text{D}}(\mathbf{k})=M_{d+2}\gamma_{d+2}+M_{d+1}\gamma_{d+1}+\sum^{d}_{j=1}k_j\gamma_j$ with $M_{d+1}=\tilde{m}_{d+1}-\sum^{d}_{j=1}k^2_j$ and $M_{d+2}=\tilde{m}_{d+2}-\sum^{d-1}_{j=1}k^2_j$.
Compared with the 1st-order topological phase, $H_{d\text{D}}(\mathbf{k})$ is now rewritten as
\begin{equation}\label{eqs5}
H_{d\text{D}}(k_1,\cdots,k_{d-1},-\mathrm{i}\partial_{r_d})=\gamma_{d+1}\left(M_{d+1}-\mathrm{i}\gamma_{d+1}\gamma_d\frac{\partial}{\partial_{r_d}}\right)+\sum^{d-1}_{j=1}k_j\gamma_j+M_{d+2}\gamma_{d+2}
\end{equation}
with $M_{d+1}=\tilde{m}_{d+1}+{\partial^2}/{\partial^2_{r_d}}-\sum^{d-1}_{j=1}k^2_j$ and $M_{d+2}=\tilde{m}_{d+2}-\sum^{d-1}_{j=1}k^2_j$ for $k_d$-OBC.
However, for $k_{d'}$-OBC with an arbitrary $d'\neq d$ ($d'=1,2,\cdots,d-1$),
$H_{d\text{D}}(\mathbf{k})$ is rewritten as
\begin{equation}\label{eqs6}
H_{d\text{D}}(k_1,\cdots,k_{j\neq{d'}},\cdots,k_{d},-\mathrm{i}\partial_{r_{d'}})=\gamma_{d+1}\left(M_{d+1}-\mathrm{i}\gamma_{d+1}\gamma_{d'}\frac{\partial}{\partial_{r_{d'}}}\right)+\sum^{d}_{j=1,j\neq d'}k_j\gamma_j
+M_{d+2}\gamma_{d+2}
\end{equation}
with $M_{d+1}=\tilde{m}_{d+1}+{\partial^2}/{\partial^2_{r_{d'}}}-\sum^{d}_{j=1,j\neq d'}k^2_j$ and $M_{d+2}=\tilde{m}_{d+2}+{\partial^2}/\partial^2_{r_{d'}}-\sum^{d-1}_{j=1,j\neq d'} k^2_j$. For the Eqs.~\eqref{eqs5} and~\eqref{eqs6}, it is clear that their $M_{d+1}$-terms are same but $M_{d+2}$-terms are different. After taking the zero-energy modes $\Phi(r_d\geqslant 0)$ and $\Phi(r_{d'}\geqslant 0)$ of Eq.~\eqref{eqs3} into the above two Hamiltonians respectively, both $M_{d+1}$-term shall vanish. The remaining terms give $ H_{d\text{D}}(k_1,\cdots,k_{d-1})=\sum^{d-1}_{j=1}k_j\gamma_j+\tilde{m}_{d+2}\gamma_{d+2}$ for $k_d$-OBC but $H_{d\text{D}}(k_1,\cdots,k_{j\neq{d'}},\cdots,k_{d})=\sum^{d}_{j=1,j\neq d'}k_j\gamma_j+(\tilde{m}_{d+2}-\tilde{m}_{d-1})\gamma_{d+2}$ for $k_{d'\neq d}$-OBC, in which we have used $\Phi\frac{\partial^2}{\partial^2_{r_{d'}}}\Phi=\sum^{d}_{j=1,j\neq d'}k^2_j-\tilde{m}_{d+1}$ and ignored the $k_d^2$-term due to $\mathbf{k}\rightarrow \mathbf{0}$. One can find that all $(d-1)$D edge states can open gap, when the effective masses $\tilde{m}_{d+2}$ and $\tilde{m}_{d+2}-\tilde{m}_{d-1}$ are both non-zero.

Next we follow the previous projection approach and obtain all $(d-1)$D edge Hamiltonians
\begin{equation}\label{eqsm}
\begin{split}
&H^{\text{edge}}_{(d-1)\text{D}}(k_1,\cdots,k_{d-1})=\sum^{d-1}_{j=1}k_j\gamma^{(1)}_j+\tilde{m}_{d+2}\gamma^{(1)}_{d+2},~~~~~~~k_d-\text{OBC}, \\
H^{\text{edge}}_{(d-1)\text{D}}(k_1,\cdots,&k_{j\neq{d'}},\cdots,k_{d})=\sum^{d}_{j=1,j\neq d'}k_j\gamma^{(1)}_j
+(\tilde{m}_{d+2}-\tilde{m}_{d+1})\gamma^{(1)}_{d+2},~~~~~~~k_{d'\neq d}-\text{OBC},\\
\end{split}
\end{equation}
by using the projection operator $\mathbf{P}^{(0)}=(1-\mathrm{i}\gamma_{d+1}\gamma_{d(d')})/2$ for each low-energy $H_{d\text{D}}(\mathbf{k})$ along $k_{d(d')}$ directions. Clearly, the emergence of $(d-2)$D MDWs can be determined by observing the change of the effective mass sign for above $(d-1)$D edge Hamiltonians.
It should be noted that here $\gamma^{(1)}_{j}$ [or $\gamma^{(1)}_{d+2}$] in Eq.~\eqref{eqsm} have same forms for all edge Hamiltonians, because we can perform rotation transformation for Gamma matrices without changing the topological property of each edge Hamiltonian and finally this only makes difference for the effective mass signs. Accordingly, the topological change of these edge Hamiltonians as just the changing in the effective mass sign. That can be classified by the following three cases:
\\
\par\textbf{(i)} $\tilde{m}_{d+2}$ and all $\tilde{m}_{d+2}-\tilde{m}_{d-1}$ have same signs, which implies that there is no gapless boundary state in the $(d-2)$D real-space interfaces;
\\
\par\textbf{(ii)} The part of $\tilde{m}_{d+2}-\tilde{m}_{d-1}$ have different signs, which naturally makes their signs are different with $\tilde{m}_{d+2}$, leading to the $(d-2)$D gapless boundary states. Yet, these $(d-2)$D real-space interfaces hosted MDWs depend on the specific edge Hamiltonians with the different effective masses;
\\
\par\textbf{(iii)} All $\tilde{m}_{d+2}-\tilde{m}_{d-1}$ have same signs but that have different signs with $\tilde{m}_{d+2}$, leading to the gapless boundary states with the emergence of MDWs in the real-space $(d-2)$D interfaces perpendicular to a certain direction $r_d$.
\\
\\
For a 2nd-order topological phase obeying the Hamiltonian~\eqref{eqs4-2}, the case ii and case iii are both covered. These $(d-2)$D edge modes should hold $\tilde{m}_{d+1}=0$ and $\tilde{m}_{d+2}-\tilde{m}_{d+1}=0$ simultaneously, i.e., $M_{d+1}=M_{d+2}=0$. Similar to 1st-order topological phase, for all $(d-2)$D projective Hamiltonians, $M_{d+1}$ and $M_{d+2}$ are the results of $h_{d+1}(\mathbf{k}\rightarrow \mathbf{0})$ and $h_{d+2}(\mathbf{k}\rightarrow \mathbf{0})$ respectively, which implies that these MDWs on the $(d-2)$D real-space interfaces correspond to the $(d-2)$D momentum-space surfaces where both $h_{d+1}(\mathbf{k})$ and $h_{d+2}(\mathbf{k})$ vanishes, defined the 2nd-order BISs $\mathcal{B}_2\equiv \{\mathbf{k}|h_{d+1}=h_{d+2}=0\}$. This gives a MDW-BIS duality of 2nd-order topological phases, showing that the existence of $(d-2)$D gapless boundary states uniquely corresponds to the emergence of $2$nd-order BIS [see Fig.~\ref{Figs1}].
It is worth mentioning that we hereby do not emphasize the crystalline symmetries of the Hamiltonian (\ref{eqs4-2}), because the above proof is general. On the contrary, we focus on a lattice Hamiltonian can satisfy the above requirements which leads to the MDWs in Eq.~(\ref{eqsm}). And then, no matter what kinds of crystalline symmetries it has, we can always use the $2$nd-order (or higher-order) BISs to characterize the $2$nd-order (or higher-order) topological phases in momentum space.

For the simplest case iii, we only use the $(d-1)$D edge Hamiltonian (\ref{eqsm}) of $k_d$-OBC to characterize the 2nd-order topological properties of bulk Hamiltonian, since there is no MDW in the $(d-2)$D interfaces along the $r_d$ direction. We regard the $(d-1)$D edge Hamiltonian of $k_d$-OBC as a subsystem and further obtain its $(d-2)$D edge Hamiltonian
\begin{equation}\label{eq2d}
\begin{split}
H^{\text{edge}}_{(d-2)\text{D}}(k_1,\cdots,k_{d-2})=\sum^{d-2}_{j=1}k_j\gamma^{(2)}_j
\end{split}
\end{equation}
by using the projection operator $\mathbf{P}^{(1)}=(1-\mathrm{i}\gamma^{(1)}_{d+2}\gamma^{(1)}_{d-1})/2$ in the eigenspace of $\mathrm{i}\gamma^{(1)}_{d+2}\gamma^{(1)}_{d-1}=-1$. Namely, we have $H^{\text{edge}}_{(d-2)\text{D}}=\mathbf{P}^{(1)^{\dagger}}\mathbf{P}^{(0)^{\dagger}}H_{d\text{D}}(\mathbf{k})\mathbf{P}^{(0)}\mathbf{P}^{(1)}$. Note that $H^{\text{edge}}_{(d-2)\text{D}}$ are topologically equivalent to project the $(d-1)$D projective Hamiltonian
\begin{equation}\label{eq2f}
\mathcal{H}^{\text{proj}}_{(d-1)\text{D}}(\mathbf{k}^{(1)})=\sum^{d-1}_{j=1}h_j\gamma^{(1)}_j+h_{d+2}\gamma^{(1)}_{d+2}
\end{equation}
along $k_{d-1}$ in momentum space, i.e., the $(d-2)$D projective Hamiltonian
\begin{equation}
\mathcal{H}^{\text{proj}}_{(d-2)\text{D}}(\mathbf{k}^{(2)})=\sum^{d-2}_{j=1}h_j\gamma^{(2)}_j.
\end{equation}
Considering that $\mathcal{H}^{\text{proj}}_{(d-1)\text{D}}(\mathbf{k}^{(1)})$ describes a subsystem which is similar to the 1st-order topological phases, thus its topology can be treated as an integer invariant on the projective $s$th-order BISs
$\mathcal{B}^{(1)}_{\text{proj},s}\equiv \{\mathbf{k}^{(1)}|h_{d+2}=h_{d-1}=\cdots=0\}$, say $w_2$.

We next give the topological number $\mathcal{V}_2$ to characterize the bulk Hamiltonian~\eqref{eqs4-2} through the two topological indexes $\mathcal{V}_1\equiv w_1$ and $w_2$. Since the construction of the 2nd-order topological phases is based on the 1st-order topological phases with the nonzero $w_1$, i.e., the existence of the $(d-1)$D edge states, the extra mass term $h_{d+2}(\mathbf{k})$ can open energy gap of $(d-1)$D edge states and whose dispersion is also destroyed~\cite{mong2011edge-s,kunst2017anatomy-s}. If the system before adding the $h_{d+2}(\mathbf{k})$-term is trivial ($\mathcal{V}_1=0$) and then there is no $(d-1)$D edge states, the 2nd-order topological phases should be trivial even if adding the additional mass term. Hence we take $\text{sgn}(|\mathcal{V}_1|)$ to characterize the topology of $(d-1)$D edge states [or say the existence of $(d-1)$D edge states], where $\text{sgn}(|w_1|)=1$ is for topological and $\text{sgn}(|\mathcal{V}_1|)=0$ is for trivial. Finally, for the Hamiltonian (\ref{eqs4-2}), these $(d-2)$D edge states should inherit the topological properties of $(d-1)$D edge states and characterized by
\begin{equation}
\mathcal{V}_2=\text{sgn}(|\mathcal{V}_1|)w_2=\text{sgn}(|w_1|)w_2.
\end{equation}
On the other hand, for $k_dk_{d-1}$-OBCs, the two components of $h_{d+2}(\mathbf{k}^{(2)})$ and $h_{d-1}(k_{d-1})$ further vanish in this projection process of $\mathbf{P}^{(1)}$. The $(d-2)$D edge states on the $(d-2)$D interfaces perpendicular to both $r_d$ and $r_{d-1}$ directions are only characterized by $\text{BZ}^{(2)}$ with $\mathbf{k}^{(2)}=(k_{1},\cdots,k_{d-2})$, which is the projection of the projective 1st-order BISs $\mathcal{B}^{(1)}_{\text{proj},1}\equiv \{\mathbf{k}^{(1)}|h_{d+2}=0\}$ on the surfaces of $h_{d-1}(k_{d-1})=0$ and has the boundaries $\mathcal{B}^{(1)}_{\text{proj},2}\equiv \{\mathbf{k}^{(1)}|h_{d+2}=h_{d-1}=0\}$ [see Fig.~\ref{Figs1}(e2)]. Particularly, the vanishing $h_{d+1}(\mathbf{k})$-term and $h_{d+2}(\mathbf{k})$-term gives $\mathcal{B}_2$, which is equivalent the union of $\mathcal{B}^{(0)}_{\text{proj},1}$ and $\mathcal{B}^{(1)}_{\text{proj},1}$, i.e., $\mathcal{B}_2\Leftrightarrow\mathcal{B}^{(0)}_{\text{proj},1}\cup \mathcal{B}^{(1)}_{\text{proj},1}$. This implies that $\mathcal{B}_2$ shall generate $\mathcal{B}^{(1)}_{\text{proj},1}$ with the projection of $\mathcal{B}^{(0)}_{\text{proj},1}$ at the surface of $h_d=0$. Hence $\mathcal{B}^{(1)}_{\text{proj},1}$ reflects the existence of $\mathcal{B}_2$, which allows us to identify $2$nd-order topological phases though $\mathcal{B}^{(1)}_{\text{proj},1}$ in $(d-1)$D momentum subspace $\mathbf{k}^{(1)}$ under the MDW-BIS duality. For the case ii, the establishment of the topological characterization is similar to the case iii, except that we need to confirm which $(d-1)$D projective Hamiltonian captures these $(d-1)$D gapless boundary states. Here we will not discuss it in detail.

\subsubsection{3. $3$rd-order topological phases}

We add a mass term $h_{d+3}(k_1,\cdots,k_{d-2})$ in the Hamiltonian (\ref{eqs4-2}) and obtain
\begin{equation}\label{eqs3d}
\mathcal{H}_{d\text{D}}(\mathbf{k})=\sum^{d}_{j=1}h_j\gamma_j+h_{d+1}\gamma_{d+1}+h_{d+2}\gamma_{d+2}+h_{d+3}\gamma_{d+3},
\end{equation}
where the Gamma matrices still satisfy the Clifford algebra. This Hamiltonian~\eqref{eqs3d} can describe the $d$D $3$rd-order topological phases. Similar to the 1st-order and 2nd-order topological phases, we can obtain the duality between the MDWs in $(d-3)$D real-space interfaces and 3rd-order BISs [see Fig.~\ref{Figs1}]. Firstly, we regard these $(d-1)$D edge Hamiltonians $H^{\text{edge}}_{(d-1)\text{D}}$ of $k_d$-OBC and $k_{d'}$-OBCs as the $(d-1)$D subsystems in Eq.~(\ref{eqsm}) and obtain their zero-energy modes of $(d-2)$D edges, which is similar to the 1st-order case. After performing the straightforward calculation, the $(d-2)$D edge Hamiltonians reads
\begin{equation}\label{eqsms}
\begin{split}
&H^{\text{edge}}_{(d-2)\text{D}}(k_1,\cdots,k_{d-2})=\sum^{d-2}_{j=1}k_j\gamma^{(2)}_j+\tilde{m}_{d+3}\gamma^{(2)}_{d+3},~~~~k_dk_{d-1}-\text{OBCs}, \\
H^{\text{edge}}_{(d-2)\text{D}}(&k_1,\cdots,k_{j\neq{d''}},\cdots,k_{d-1})=\sum^{d-1}_{j=1,j\neq d''}k_j\gamma^{(2)}_j
+(\tilde{m}_{d+3}-\tilde{m}_{d+2})\gamma^{(2)}_{d+3},~~~~k_dk_{{d''}\neq {d-1}}-\text{OBCs},\\
H^{\text{edge}}_{(d-2)\text{D}}(k_1,\cdots,&k_{j\neq{d'},{d''}},\cdots,k_{d})=\sum^{d}_{j=1,j\neq d',d''}k_j\gamma^{(2)}_j
+(\tilde{m}_{d+3}-\tilde{m}_{d+2}+\tilde{m}_{d+1})\gamma^{(2)}_{d+3},~k_{d'''\neq d', d''}k_{j}-\text{OBCs}
\end{split}
\end{equation}
by taking OBCs along two arbitrary directions, where $d'=1,2,\cdots, d-1$ and $d''=1,2,\cdots, d-2$. Similarly, the signs of effective masses still have three cases as follows:
\\
\par \textbf{(i)} $\tilde{m}_{d+3}$, all $\tilde{m}_{d+3}-\tilde{m}_{d+2}$, and all $\tilde{m}_{d+3}-\tilde{m}_{d+2}+\tilde{m}_{d+1}$ have same signs, leading to no gapless boundary state in the $(d-3)$D real-space interfaces;
\\
\par \textbf{(ii)} Some of $\tilde{m}_{d+3}-\tilde{m}_{d+2}$ or some of $\tilde{m}_{d+3}-\tilde{m}_{d+2}+\tilde{m}_{d+1}$ have different signs, which naturally makes them have different signs with the remaining effective masses, leading to the $(d-3)$D gapless boundary states. But these $(d-3)$D real-space interface hosted MDWs depend on the specific edge Hamiltonians with the different effective masses;
\\
\par \textbf{(iii)} All $\tilde{m}_{d+3}-\tilde{m}_{d+2}$ and all $\tilde{m}_{d+3}-\tilde{m}_{d+2}+\tilde{m}_{d+1}$ have same signs but have different signs with $\tilde{m}_{d+3}$, leading to the $(d-3)$D gapless boundary states due to emerging MDWs in the real-space $(d-3)$D interfaces perpendicular to the certain direction $r_d$.
\\
\par For a $3$rd-order topological phase obeying the Hamiltonian~\eqref{eqs3d}, both cases ii and iii are covered. These $(d-3)$D MDWs should hold $\tilde{m}_{d+3}=0$, $\tilde{m}_{d+3}-\tilde{m}_{d+2}=0$, and $\tilde{m}_{d+3}-\tilde{m}_{d+2}+\tilde{m}_{d+1}=0$ simultaneously, i.e., $M_{d+1}=M_{d+2}=M_{d+3}=0$. This implies that there are momentum-space 3rd-order BISs $\mathcal{B}_3\equiv \{\mathbf{k}|h_{d+1}=h_{d+2}=h_{d+3}=0\}\Leftrightarrow\mathcal{B}^{(0)}_{\text{proj},1}\cup \mathcal{B}^{(1)}_{\text{proj},1}\cup \mathcal{B}^{(2)}_{\text{proj},1}$. Especially for the case iii, the topology of $(d-2)$D projective Hamiltonian
\begin{equation}\label{eq3d}
\mathcal{H}^{\text{proj}}_{(d-2)\text{D}}(\mathbf{k}^{(2)})=\sum^{d-2}_{j=1}h_j\gamma^{(2)}_j+h_{d+3}\gamma^{(2)}_{d+3}
\end{equation}
can be treated as an integer invariant on the projective $s$th-order BISs $\mathcal{B}^{(2)}_{\text{proj},s}\equiv \{\mathbf{k}^{(2)}|h_{d+3}=h_{d-2}=\cdots=0\}$, say $w_3$. For the bulk Hamiltonian (\ref{eqs3d}), hence the $(d-3)$D edge states on the real-space interfaces perpendicular to $r_dr_{d-1}$ directions are characterized by $\mathcal{V}_3=\text{sgn}(|\mathcal{V}_2|)w_3=\text{sgn}(|w_1w_2|)w_3$.

\subsubsection{4. $n$th-order topological phases}

By repeating the above process of adding mass terms until there are $n$ mass terms in the bulk Hamiltonian as
\begin{equation}\label{eq1-sm}
\begin{split}
\mathcal{H}_{d\text{D}}(\mathbf{k})=\mathcal{H}_{\mathbf{k}}=\sum^{d}_{j=1}h_j(k_j)\gamma_j+\sum^{n}_{l=1}h_{d+l}(k_1,\cdots,k_{d-l+1})\gamma_{d+l},
\end{split}
\end{equation}
we can take $k_d k_{d-1}\cdots k_{d-n+1}$-OBCs and obtain the $(d-n)$D edge Hamiltonian as
\begin{equation}
H^{\text{edge}}_{(d-n)\text{D}}(k_1,\cdots,k_{d-n})=\sum^{d-n}_{j=1}k_j\gamma^{(n)}_j,
\end{equation}
which holds the zero-energy modes on all $(d-n)$D edges perpendicular to $r_dr_{d-1}\cdots r_{d-n+1}$ directions. Further, we can find a duality between the MDWs and the $n$th-order BISs
\begin{equation}\label{eqs9}
(d-n)\text{D MDWs}\; \Longleftrightarrow\; \mathcal{B}_n\equiv\{\mathbf{k}|h_{d+1}=\cdots =h_{d+n}=0\}\Leftrightarrow \mathcal{B}^{(0)}_{\text{proj},1}\cup \mathcal{B}^{(1)}_{\text{proj},1}\cup\cdots\cup\mathcal{B}^{(n-1)}_{\text{proj},1}.
\end{equation}
The existence of MDWs within $(d-n)$D crossing real-space interfaces gives the $(d-n)$D gapless edge states, which corresponds to the emergence of $\mathcal{B}_n$ in original momentum space.

Next we characterize the $n$th-order topology similar to the previous case iii. By using the projection operator $\mathbf{P}^{(n-1)}=(1-\mathrm{i}\gamma^{(n-1)}_{d+n}\gamma^{(n-1)}_{d-n+1})/2$ in the eigenspace of $\mathrm{i}\gamma^{(n-1)}_{d+n}\gamma^{(n-1)}_{d-n+1}=-1$ along $k_{d-n+1}$ to the $(d-n+1)$D projective Hamiltonian
\begin{equation}\label{eq4f}
\mathcal{H}^{\text{proj}}_{(d-n+1)\text{D}}(\mathbf{k}^{(n-1)})=\sum^{d-n+1}_{j=1}h_j\gamma^{(n-1)}_j+h_{d+n}\gamma^{(n-1)}_{d+n},
\end{equation}
the above $(d-n)$D edge Hamiltonian are topologically equivalent to
\begin{equation}
\mathcal{H}^{\text{proj}}_{(d-n)\text{D}}(\mathbf{k}^{(n)})=\sum^{d-n}_{j=1}h_j\gamma^{(n)}_j.
\end{equation}
Thus the $(d-n)$D edge states only exist the projections of the $n$th-order BISs $\mathcal{B}_n$ on the interfaces of $h_{d-n+1}(k_{d-n+1})=0$. This defines the effective Brillouin zone $\text{BZ}^{(n)}$ with the momenta $\mathbf{k}^{(n)}=(k_{1},\cdots,k_{d-n})$ and the $\text{BZ}^{(n)}$ boundaries are determined by the projective 2nd-order BISs $\mathcal{B}^{(n-1)}_{\text{proj},2}\equiv \{\mathbf{k}^{(n-1)}|h_{d+n}=h_{d-n+1}=0\}$ [see Fig.~\ref{Figs1}]. The topology of $\mathcal{H}^{\text{proj}}_{(d-n+1)\text{D}}$ can be treated as an integer invariant on the projective $s$th-order BISs $\mathcal{B}^{(n-1)}_{\text{proj},s}=\{\mathbf{k}^{(n-1)}|h_{d+n}=h_{d-n+1}=\cdots=0\}$, say $w_n$. Finally, for the bulk Hamiltonian~\eqref{eq1-sm}, the zero-energy modes on all $(d-n)$D edges perpendicular to $r_dr_{d-1}\cdots r_{d-n+2}$ directions are characterized by
\begin{equation}
\mathcal{V}_n=\text{sgn}(|\mathcal{V}_{n-1}|)w_n=\text{sgn}\left(|w_1w_2\cdots w_{n-1}|\right)w_{n}.
\end{equation}
Note that $w_i$ describes a winding or Chern number of the projective Hamiltonian $\mathcal{H}^{\text{proj}}_{(d-i+1)D}(\mathbf{k}^{(i-1)})$, thus here each $w_i$ is classified by $\mathbbm{Z}$ invariant and the $n$th-order topology of bulk Hamiltonian is also classified by the integer invariant. Remarkably, if the topology of the projective Hamiltonian $\mathcal{H}^{\text{proj}}_{(d-i+1)D}(\mathbf{k}^{(i-1)})$ for $i=1,\cdots,n-1$ is $\mathbbm{Z}_2$-classified but it can be determined by $\text{sgn}(w_i)$ where $w_i$ characterizes a $\mathbbm{Z}$-classified Hamiltonian which keeps the dimensions and $h$-components be same as the $\mathbbm{Z}_2$-classified Hamiltonian but changes the symmetries (i.e., keeping the dimensions but changing the symmetries result in $\mathbbm{Z}_2\rightarrow \mathbbm{Z}$), our characterization theory still applies but now $\mathcal{H}^{\text{proj}}_{(d-i+1)D}(\mathbf{k}^{(i-1)})$ is needed to be replaced by the $\mathbbm{Z}$-classified Hamiltonian. Accordingly, the index $w_i$ can be obtained by the topological charges of this $\mathbbm{Z}$-classified Hamiltonian and the topological number $\text{sgn}|w_{i}|$ for the $\mathbbm{Z}_2$-classified Hamiltonian is naturally recovered in $\mathcal{V}_n$. As we show that in the below dynamical characterization, the former case can be observed in $3$D $2$nd-order TI, while the later special case can be observed in $2$D $2$nd-order and $3$D $3$rd-order TIs.

\subsubsection{5. Applications of MDW-BIS duality in three typical models}

\begin{figure}[!htbp]
\begin{center}
\includegraphics[width=\columnwidth]{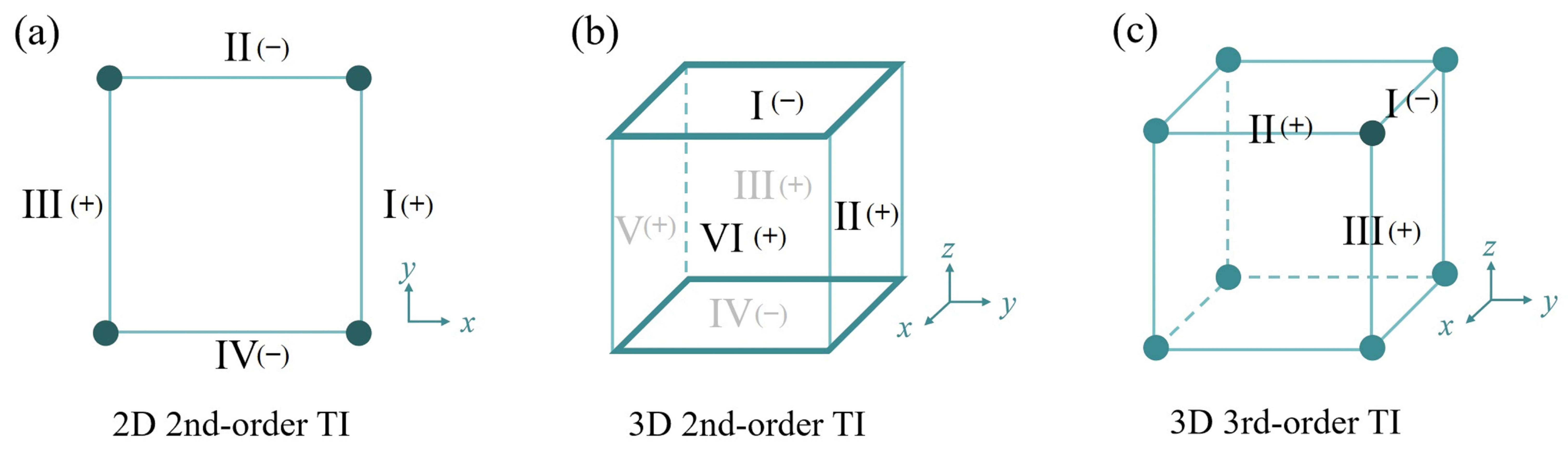}
\caption{Sketches of $0$D corner states for 2D and 3D system and $1$D hinge states in 3D system.
Four edges and whose mass sign are I(+), II(-), III(+), and IV(-) in (a), where the signs of mass term alternately change along the counterclockwise and hold four corner states in 2D 2nd-order TI.
Six surfaces and whose mass sign are I(-), II(+), III(+), IV(+), V(+), and VI(-) in (b), where the hinges along $x$ and $y$ hold MDWs due to the sign change of mass terms in 3D 2nd-order TI.
In 3D 3rd-order TI of (c), the changing sign of mass terms only are along the hinges of $x$ and $y$, which products eight corner states.}
\label{Figs2}
\end{center}
\end{figure}

We first focus on a 2D 2nd-order TI which can be constructed by copying the quantum anomalous hall insulator and adding a mass term to break its time-reversal symmetry~\cite{liu2013detecting-s}. The corresponding Hamiltonian is
\begin{equation}\label{eqs10}
\begin{split}
&\mathcal{H}_{2\text{D}}(k_x,k_y)=h_1\gamma_1+h_2\gamma_2+h_3\gamma_3+h_4\gamma_4,\\
h_{1,2}=t_{\text{so}}\sin k_{x,y},&~h_3=m_3-t_0(\cos k_x+\cos k_y),~
h_4=m_4-t_0\cos k_x,
\end{split}
\end{equation}
with $\gamma_1=\tau_z\otimes\sigma_x$, $\gamma_2=\tau_z\otimes\sigma_y$, $\gamma_3=\tau_z\otimes\sigma_z$, and $\gamma_4=\tau_y\otimes\sigma_0$, where $\sigma_{x,y,z}$ and $\tau_{x,y,z}$ are both Pauli matrices. $\tau_0$ and $\sigma_0$ are identity matrices.
This 2D Hamiltonian only has chiral symmetry $U_c^{\dagger}\mathcal{H}_{\mathbf{k}}U_c=-\mathcal{H}_{\mathbf{k}}$ with $U_c=\tau_x\otimes\sigma_0$
and belongs to the class AIII.
The system hosts inversion symmetry $\mathcal{I}^{\dagger}\mathcal{H}_{\mathbf{k}}\mathcal{I}=\mathcal{H}_{-\mathbf{k}}$ with
$\mathcal{I}=\tau_0\otimes\sigma_z$, $x$-direction reflection symmetry
$\mathcal{R}_x^{\dagger}\mathcal{H}_{k_x,k_y}\mathcal{R}_x=\mathcal{H}_{-k_x,k_y}$ with
${\mathcal{R}_x}=\tau_y\otimes\sigma_x$ and $C_4$-rotation symmetry
$C_4^{\dagger}\mathcal{H}_{k_x,k_y}C_4=\mathcal{H}_{-k_y,k_x}$
with $C_4=\tau_0\otimes\text{e}^{-\mathrm{i}\frac{\pi}{4}\sigma_z}$.
The existence of 2nd-order BISs $\mathcal{B}_2$ can easily given by $|m_4|<t_0$ and $|m_3-m_4|<t_0$.
Under the approximation of $(k_x,k_y)\rightarrow (0,0)$, we project the low-energy Hamiltonian of $\mathcal{H}_{2\text{D}}$ into the edges of $x$ and $y$ by using $\mathbf{P}^{(0)}=(1-\mathrm{i}\gamma_{3}\gamma_{2})/2$
and $\mathbf{P}^{(0)}=(1-\mathrm{i}\gamma_{3}\gamma_{1})/2$ respectively. The Hamiltonians at four edges [$\text{I}(x>0),\text{II}(y>0),\text{III}(x<0),\text{IV}(y<0)$] are written as
\begin{equation}\label{eqs11}
\begin{split}
&H^{\text{edge}}_\text{I}(k_x)=t_{\text{so}}k_x\sigma_z-\left(m_4-t_0\right)\sigma_y,\\
&H^{\text{edge}}_\text{II}(k_y)=-t_{\text{so}}k_y\sigma_z+\left({m_3-m_4}-{t_0}\right)\sigma_y,\\
&H^{\text{edge}}_\text{III}(k_x)=-t_{\text{so}}k_x\sigma_z-\left(m_4-t_0\right)\sigma_y,\\
&H^{\text{edge}}_\text{IV}(k_y)=t_{\text{so}}k_y\sigma_z+\left({m_3-m_4}-{t_0}\right)\sigma_y.\\
\end{split}
\end{equation}
One can find that all the terms of brackets ($\cdots$) have the same sign under the parameter conditions of $|m_4|<t_0$ and $|m_3-m_4|<t_0$ (i.e. the existence of $\mathcal{B}_2$). The signs of last terms in $H^{\text{edge}}_\text{I},H^{\text{edge}}_\text{II},H^{\text{edge}}_\text{III}$, and $H^{\text{edge}}_\text{IV}$ alternately change along the counterclockwise of edges, which implies that the MDWs should be produced at four corners and each corner holds a zero-energy mode [see Fig.~\ref{Figs2}(a) and Fig.~\ref{Figs3}(g)]. Accordingly, we clearly observe that the MDWs of four corners are corresponding to the 2nd-order BISs $\mathcal{B}_2$, which are four points in Fig.~\ref{Figs3}(d).

Further, we consider the 3D 2nd-order TI of main text and write the corresponding Hamiltonian again
\begin{equation}\label{eqs12}
\begin{split}
&\mathcal{H}_{3\text{D}}(k_x,k_y,k_z)=h_1\gamma_1+h_2\gamma_2+h_3\gamma_3+h_4\gamma_4+h_5\gamma_5,\\
h_{1,2,3}=t_{\text{so}}\sin k_{x,y,z},~
&h_4=m_4-t_0(\cos k_x+\cos k_y+\cos k_z),~
h_5=m_5-t_0(\cos k_x+\cos k_y),
\end{split}
\end{equation}
with $k_{1,2,3}=k_{x,y,z}$. The Gamma matrices are taken as $\gamma_1=\sigma_x\otimes \tau_0$, $\gamma_2=\sigma_y\otimes \tau_0$, $\gamma_3=\sigma_z\otimes \tau_x$, $\gamma_4=\sigma_z\otimes \tau_z$, and $\gamma_5=\sigma_z\otimes \tau_y$.
This 3D Hamiltonian only has particle-hole symmetry
$U_p^{\dagger}\mathcal{H}^{*}_{-\mathbf{k}}U_p=-\mathcal{H}_{\mathbf{k}}$ with $U_p=\sigma_x\otimes \tau_z$ and belongs to the class D. The system hosts $C^{z}_4$-rotation symmetry $C_4=\tau_0\otimes\text{e}^{-\mathrm{i}\frac{\pi}{4}\sigma_z}$ and anti-reflection symmetry $\bar{\mathcal{R}}_j^{\dagger}\mathcal{H}_{k_j}\bar{\mathcal{R}}_j=-\mathcal{H}_{-k_j}$ along three direction of $j=x,y,z$, with $\bar{\mathcal{R}}_x=\sigma_x\otimes \tau_0$, $\bar{\mathcal{R}}_y=\sigma_y\otimes \tau_0$, and $\bar{\mathcal{R}}_z=\sigma_z\otimes \tau_x$.
Firstly, we can easily obtain the parameters of the existence of 2nd-order BISs $\mathcal{B}_2$ are $|m_5|<2t_0$ and $|m_4-m_5|<t_0$.
Secondly, we project the low-energy Hamiltonian of $\mathcal{H}_{3\text{D}}$ into the surfaces of $xy$, $yz$ and $xz$ by using $\mathbf{P}^{(0)}=(1-\mathrm{i}\gamma_{4}\gamma_{3})/2$, $\mathbf{P}^{(0)}=(1-\mathrm{i}\gamma_{4}\gamma_{1})/2$, and $\mathbf{P}^{(0)}=(1-\mathrm{i}\gamma_{4}\gamma_{2})/2$ respectively. The Hamiltonians of six 2D surfaces [$\text{I}(z>0),\text{II}(x>0),\text{III}(y>0),\text{IV}(z<0),\text{V}(x<0),\text{VI}(y<0)$] are written as
\begin{equation}\label{eqs13}
\begin{split}
&H^{\text{surf}}_\text{I}(k_x,k_y)=t_{\text{so}}k_x\sigma_z+t_{\text{so}}k_y\sigma_x+\left(m_5-2t_0\right)\sigma_y,\\
&H^{\text{surf}}_\text{II}(k_y,k_z)=-t_{\text{so}}k_y\sigma_z-t_{\text{so}}k_z\sigma_x-\left(m_4-m_5-t_0\right)\sigma_y,\\
&H^{\text{surf}}_\text{III}(k_x,k_z)=t_{\text{so}}k_x\sigma_z-t_{\text{so}}k_z\sigma_x-\left({m_4-m_5}-{t_0}\right)\sigma_y,\\
&H^{\text{surf}}_\text{IV}(k_x,k_y)=-t_{\text{so}}k_x\sigma_z-t_{\text{so}}k_y\sigma_x+\left(m_5-2t_0\right)\sigma_y,\\
&H^{\text{surf}}_\text{V}(k_y,k_z)=t_{\text{so}}k_y\sigma_z+t_{\text{so}}k_z\sigma_x-\left(m_4-m_5-t_0\right)\sigma_y,\\
&H^{\text{surf}}_\text{VI}(k_x,k_z)=-t_{\text{so}}k_x\sigma_z+t_{\text{so}}k_z\sigma_x-\left({m_4-m_5}-{t_0}\right)\sigma_y,\\
\end{split}
\end{equation}
One can find that the terms of brackets ($\cdots$) in $H^{\text{surf}}_{\text{II},\text{III}}$ and $H^{\text{surf}}_{\text{V},\text{VI}}$ are exactly same, thus the mass terms do not change sign no matter what parameters of $m_4$ and $m_5$. Naturally, there is no MDW in the hinges along $z$-axis. However, under the parameter conditions of $|m_5|<2t_0$ and $|m_4-m_5|<t_0$ (i.e. the existence $\mathcal{B}_2$), the last terms of $H^{\text{surf}}_{\text{I},\text{IV}}$ have different signs with the remaining Hamiltonians, and the mass terms change signs on the two sides of $x$ and $y$ edges, which induces the MDWs [see Fig.~\ref{Figs2}(b)]. These MDWs on the hinges along $x$ and $y$ are corresponding to the 2nd-order BISs $\mathcal{B}_2$ (i.e., $h_4=h_5=0$) which presents two rings in the inset of Fig.~3(a) of the main text.

Finally, we consider a 3D 3rd-order TI and the corresponding Hamiltonian is
\begin{equation}\label{eqs14}
\begin{split}
&\mathcal{H}_{3\text{D}}
(k_x,k_y,k_z)=h_1\gamma_1+h_2\gamma_2+h_3\gamma_3+h_4\gamma_4+h_5\gamma_5+h_6\gamma_6,\\
&h_{1,2,3}=t_{\text{so}}\sin k_{x,y,z},~
h_4=t_0(\cos k_x+\cos k_y+\cos k_z)-m_4,~\\
&h_5=t_0(\cos k_x+\cos k_y)-m_5,~h_6=t_0\cos k_x-m_6,
\end{split}
\end{equation}
with $k_{1,2,3}=k_{x,y,z}$. The Gamma matrices are taken as $\gamma_1=s_y\otimes\sigma_y\otimes\tau_z$, $\gamma_2=s_y\otimes\sigma_z\otimes\tau_0$, $\gamma_3=s_z\otimes\sigma_0\otimes\tau_0$, $\gamma_4=s_x\otimes\sigma_0\otimes\tau_0$, $\gamma_5=s_y\otimes\sigma_x\otimes\tau_0$ and $\gamma_6=s_y\otimes\sigma_y\otimes\tau_x$, where $s_{x,y,z}$, $\sigma_{x,y,z}$ and $\tau_{x,y,z}$ are both Pauli matrices. $s_0$, $\sigma_0$, and $\tau_0$ are identify matrices.
This 3D Hamiltonian only has chiral symmetry with $U_c=s_y\otimes\sigma_y\otimes\tau_y$ and belongs to the class AIII. The systems hosts the inversion symmetry $\mathcal{I}=s_x\otimes\sigma_z\otimes\tau_x$ and reflection symmetry $\mathcal{R}_x=s_0\otimes\sigma_0\otimes\tau_x$,
$\mathcal{R}_y=s_0\otimes\sigma_x\otimes\tau_y$,
$\mathcal{R}_z=s_x\otimes\sigma_y\otimes\tau_y$ along $x,y,z$-direction.
We can easily obtain the parameters of the existence of $\mathcal{B}_3$ are $|m_6|<t_0$, $|m_4-m_5|<t_0$ and $|m_5-m_6|<t_0$.
Then we project the low-energy model of $\mathcal{H}_{3\text{D}}$ into the hinges of $x$, $y$ and $z$ by using $\mathbf{P}^{(0)}=(1-\mathrm{i}\gamma_{4}\gamma_{3})/2$ and $\mathbf{P}^{(1)}=(1-\mathrm{i}\gamma^{(1)}_{5}\gamma^{(1)}_{2})/2$ for $k_zk_y$-OBCs, $\mathbf{P}^{(0)}=(1-\mathrm{i}\gamma_{4}\gamma_{3})/2$ and $\mathbf{P}^{(1)}=(1-\mathrm{i}\gamma^{(1)}_{5}\gamma^{(1)}_{1})/2$ for $k_zk_x$-OBCs, and $\mathbf{P}^{(0)}=(1-\mathrm{i}\gamma_{4}\gamma_{1})/2$ and $\mathbf{P}^{(1)}=(1-\mathrm{i}\gamma^{(1)}_{5}\gamma^{(1)}_{2})/2$ for $k_xk_y$-OBCs, respectively.
The Hamiltonians of three hinges [$\text{I}(y>0,z>0),\text{II}(x>0,z>0),\text{III}(x>0,y>0)$] are written as
\begin{equation}\label{eqs15}
\begin{split}
&H^{\text{hinge}}_\text{I}(k_x)=t_{\text{so}}k_x\sigma_z-\left[t_0-m_6\right]\sigma_y,\\
&H^{\text{hinge}}_\text{II}(k_y)=-\frac{t_{\text{so}}}{2}k_y\sigma_z+\frac{\left[t_0-(m_5-m_6)\right]}{2}\sigma_y,\\
&H^{\text{hinge}}_\text{III}(k_z)=-\frac{t_{\text{so}}}{2}k_z\sigma_z+\frac{\left[(t_0-m_6)+t_0-(m_4-m_5)\right]}{2}\sigma_y,\\
\end{split}
\end{equation}
One can find that the terms of brackets [$\cdots$] in $H^{\text{hinge}}_{\text{I},\text{II},\text{III}}$ have the same signs under the parameter conditions with the existence $\mathcal{B}_3$. On the corners of the crossing directions of $x$, $y$, and $z$, we observe that these mass terms change signs in Fig.~\ref{Figs2}(c), which induces the MDWs. When we consider all hinge Hamiltonians, these MDWs on eight corners are corresponding to the 3nd-order BISs $\mathcal{B}_3$ (i.e., $h_4=h_5=h_6=0$) which are eight green points in Fig.~\ref{Figs1}(a7).

\subsection{II. Higher-order topological phase transitions}

From above dimensional reduction with $i=1,2,\cdots,n$ to construct the $n$th-order topological phases, we can treat the $(d-i+1)$D gapless boundary modes as a massless Dirac system, and then the MDWs of the $i$th-order topological phases are indeed the boundary states of the projective Hamiltonian like Eq.~\eqref{eqs1}, Eq.~\eqref{eq2f}, Eq.~\eqref{eq3d}, and Eq.~\eqref{eq4f}, giving an effective 1st-order $(d-i+1)$D gapped subsystem
\begin{equation}\label{sh2}
\mathcal{H}_{\mathbf{k}^{(i-1)}}\equiv \mathcal{H}^{\text{proj}}_{(d-i+1)D}(\mathbf{k}^{(i-1)})=\sum_{j\in D^{(i-1)}}h_j(k_j)\gamma^{(i-1)}_j+h_{d+i}(\mathbf{k}^{(i-1)})\gamma^{(i-1)}_{d+i}
\end{equation}
with $D^{(i-1)}$ being a subset of $\{1,2,\dots,d\}$ with $(d-i+1)$ elements. Here the first term captures the boundary states of the $(i-1)$th-order topological phase located in the $(d-i+1)$D momentum subspace $\mathrm{BZ}^{(i-1)}\ni \mathbf{k}^{(i-1)}$, with $\mathbf{k}^{(0)}$ representing the full $d$D BZ. Accordingly, $\mathbf{k}^{(i-1)}$ denotes an effective $(d-i+1)$D BZ obtained by projecting $\mathcal{B}_{i-1}$ onto the subspace spanned by all $k_{j}$. The corresponding Gamma matrices are denoted as $\gamma^{(i-1)}_{j}$. Especially for the bulk Hamiltonian~\eqref{eq1-sm}, the $(d-n)$D boundary states are located at the real-space interfaces perpendicular to $r_d r_{d-1}\cdots r_{d-n+1}$-directions. Hence we only take $D^{(i-1)}=\{1,2,\dots,d-i+1\}$ and use the higher-order topological charges to identify the invariant $w_{i}$ for the effective $(d-i+1)$D $1$st-order phase \eqref{sh2}, as we show in the main text. It should be emphasized that the same characterization results will be obtained if another set of suitable $D^{(i-1)}$ is selected, where the principle is to choose the surface where $(d-n)$D edge states exist for dimensional reduction.

We now consider the characterization of the higher-order topological phase transitions (HOTPTs), which must be associated with the change of $w_{i}$ for one or multiple effective $1$st-order topological subsystems~\eqref{sh2}. It should be noted that the arbitrary different lower-dimensional interfaces may have gap closing, when the HOTPTs occurs. Generally, we should consider the topological changes of boundary states on all low-dimensional interfaces, i.e., taking all $D^{(i-1)}$ sets in $\mathcal{H}_{\mathbf{k}^{(i-1)}}$ and covering all lower-dimensional surfaces. As long as they have topological charges crossing the BISs, the energy gaps of these $\mathcal{H}_{\mathbf{k}^{(i-1)}}$ shall be closed, capturing the gap closing of the corresponding surfaces. Remarkably, the beauty of our characterization theory is that we still can capture the different boundary energy gap closing through the topological charges defined by $\mathcal{H}_{\mathbf{k}^{(i-1)}}$ with $D^{(i-1)}=\{1,2,\dots,d-i+1\}$. But now the behavior of charge is in addition to crossing the BISs (corresponds to the energy gap closing on the real-space interfaces denoted as $\mathcal{S}_{\parallel}$), and can also cross the border of the effective BZ, which corresponds to the gap closing perpendicular to these real-space interfaces, denoted as $\mathcal{S}_{\perp}$. The latter is equivalent to that the effective $1$st-order Hamiltonian describing $\mathcal{S}_{\perp}$ becomes gapless by the charges crossing the BISs. This result provides a simple perspective for identifying different HOTPTs theoretically and experimentally.

We next take the 3D 2nd-order TI obeying Hamiltonian~\eqref{eq1-sm} as an example to illustrate the above results. The bulk $\mathbf{h}$-vector is $\mathbf{h}_{\mathbf{k}}=(h_1,h_2,h_3,h_4,h_5)$. The $\mathbf{h}$-vector of 1D and 2D effective 1st-order topological Hamiltonian are given by $\mathbf{h}_{\mathbf{k}^{(0)}}=(h_1,h_2,h_3,h_4)$ for $\mathcal{H}_{\mathbf{k}^{(0)}}$, $\mathbf{h}_{\mathbf{k}^{(1)}}=(h_{1},h_{2},h_5)$ for $\mathcal{H}_{\mathbf{k}^{(1)}}$ corresponding to $xy$ surface Hamiltonian, $\mathbf{h}_{\mathbf{k}^{(1)}}=(h_{1},h_{3},h_5)$ for $\mathcal{H}_{\mathbf{k}^{(1)}}$ corresponding to $xz$ surface Hamiltonian, and $\mathbf{h}_{\mathbf{k}^{(1)}}=(h_{2},h_{3},h_5)$ for $\mathcal{H}_{\mathbf{k}^{(1)}}$ corresponding to $yz$ surface Hamiltonian. In this example, the type-I transition occurs when both $\mathcal{C}^{(0)}_{3,q}$ ($h_{1,3,4}=0$) and $\mathcal{C}^{(1)}_{2,q}$ ($h_{1,5}=0$) simultaneously cross $\mathcal{B}^{(0)}_{\text{proj},3}$ ($h_{2,3,4}=0$) and $\mathcal{B}^{(1)}_{\text{proj},2}$ ($h_{2,5}=0$), as the bulk energy gap closes, i.e., $h_{1,2,3,4,5}=0$ [see Fig.~\ref{Figs5}(i) D]. When only $\mathcal{C}^{(1)}_{2,q}$ ($h_{1,5}=0$) crosses $\mathcal{B}^{(1)}_{\text{proj},2}$ ($h_{2,5}=0$) or the border of $\text{BZ}^{(1)}$ ($h_{3,4}=0$), it gives $h_{1,2,5}=0$ [$xy$ surface gap close, Fig.S5 (i) E] or $h_{1,3,4,5}=0$ [$xz$ surface gap close, Fig.~\ref{Figs5}(i) B]. When $\mathcal{B}^{(1)}_{\text{proj},2}$ ($h_{2,5}=0$) cross the border of $\text{BZ}^{(1)}$ ($h_{3,4}=0$), it gives $h_{2,3,4,5}=0$ [$yz$ surface gap close, Fig.~\ref{Figs5}(i) B], where this case is actually that the charge cross the border of $\text{BZ}^{(1)}$ as we can define the topological charge with different $h$-component. The later three cases hold the type-II transitions.

We further take the 3D 3rd-order TI obeying Hamiltonian~\eqref{eq1-sm} as an example to illustrate the above results. The bulk $\mathbf{h}$-vector is $\mathbf{h}_{\mathbf{k}}=(h_1,h_2,h_3,h_4,h_5,h_6)$. The $\mathbf{h}$-vector of $xy$, $xz$, and $yz$ surface Hamiltonian are $(h_{1},h_{2},h_5,h_6)$, $(h_{1},h_{3},h_5,h_6)$, and $(h_{2},h_{3},h_5,h_6)$. The $\mathbf{h}$-vector of $x$, $y$, and $z$ hinge Hamiltonian are $(h_{1},h_6)$, $(h_{2},h_6)$, and $(h_{3},h_6)$. The $\mathbf{h}$-vector of 3D, 2D, and 1D effective $1$st-order topological Hamiltonian are given by $\mathbf{h}_{\mathbf{k}^{(0)}}=(h_1,h_2,h_3,h_4)$ for $\mathcal{H}_{\mathbf{k}^{(0)}}$, $\mathbf{h}_{\mathbf{k}^{(1)}}=(h_{1},h_{2},h_5)$ for $\mathcal{H}_{\mathbf{k}^{(1)}}$ with $\text{BZ}^{(1)}$ ($h_{3,4}=0$), and $\mathbf{h}_{\mathbf{k}^{(2)}}=(h_{1},h_6)$ for $\mathcal{H}_{\mathbf{k}^{(2)}}$ with $\text{BZ}^{(2)}$ ($h_{2,5}=0$). 
The type-I transition occurs when $\mathcal{C}^{(0)}_{3,q}$ ($h_{1,3,4}=0$) and $\mathcal{C}^{(1)}_{2,q}$ ($h_{1,5}=0$) and $\mathcal{C}^{(2)}_{1,q}$ ($h_{6}=0$) simultaneously cross $\mathcal{B}^{(0)}_{\text{proj},3}$ ($h_{2,3,4}=0$) and $\mathcal{B}^{(1)}_{\text{proj},2}$ ($h_{2,5}=0$) and $\mathcal{B}^{(2)}_{\text{proj},1}$ ($h_{1}=0$), as the bulk energy gap closes, i.e., $h_{1,2,3,4,5,6}=0$ [see Fig.~\ref{Figs7}(b4)]. When both $\mathcal{C}^{(1)}_{2,q}$ ($h_{1,5}=0$) and $\mathcal{C}^{(2)}_{1,q}$ ($h_{6}=0$) simultaneously cross $\mathcal{B}^{(1)}_{\text{proj},2}$ ($h_{2,5}=0$) and $\mathcal{B}^{(2)}_{\text{proj},1}$ ($h_{1}=0$), the $xy$ surface energy gap closes, i.e., $h_{1,2,5,6}=0$. When both $\mathcal{C}^{(1)}_{2,q}$ ($h_{1,5}=0$) and $\mathcal{C}^{(2)}_{1,q}$ ($h_{6}=0$) simultaneously cross $\text{BZ}^{(1)}$ ($h_{3,4}=0$) and $\mathcal{B}^{(2)}_{\text{proj},1}$ ($h_{1}=0$), the $xz$ surface energy gap closes, i.e., $h_{1,3,5,6}=0$ [see Fig.~\ref{Figs7}(b3)]. When $\mathcal{C}^{(2)}_{1,q}$ ($h_{6}=0$) cross $\mathcal{B}^{(2)}_{\text{proj},1}$ ($h_{1}=0$), the $x$ hinge energy gap closes, i.e., $h_{1,6}=0$. When $\mathcal{C}^{(2)}_{1,q}$ ($h_{6}=0$) cross $\text{BZ}^{(2)}$ ($h_{2,5}=0$), the $y$ hinge energy gap closes, i.e., $h_{2,6}=0$ [see Fig.~\ref{Figs7}(b2)].

\subsection{III. Dynamical characterization}

By employing the quench dynamics, each $w_i$ can be captured by the bulk Hamiltonian $\mathcal{H}_{dD}({\mathbf{k}})$. In quenching axis $\gamma_\alpha$ with $\alpha=1,2,\cdots,d+n$, we initialize a fully polarized state $\rho_\alpha$ along the opposite $\gamma_\alpha$ axis by tuning a very large constant magnetization $\delta m_\alpha$ such that $h_\alpha(\mathbf{k})\approx \delta m_\alpha \gg 0$ for $t<0$. After $t=0$, the magnetization $\delta m_\alpha$ is suddenly tuned to the topological regime, and the momentum-linked (pseudo)spin expectation $\langle \boldsymbol{\gamma}(\mathbf{k},t)\rangle$ will process around $\mathbf{h}(\mathbf{k})$. The quantum dynamics is governed by the unitary evolution operator $\mathcal{U}(t)=\text{exp}(-\mathrm{i}\mathcal{H}t)$ with the post-quenched Hamiltonian $\mathcal{H}(\mathbf{k})=\mathcal{H}_{d\text{D}}(\mathbf{k})$. We can measure the time-averaged (pseudo)spin polarization (TASP)~\cite{zhang2018dynamical-s,jia2020charge-s} of the component $\gamma_{d+1}$,
\begin{align}
    \overline{\langle\gamma_{d+1}(\mathbf{k})\rangle}_{\alpha} & \equiv\lim_{T\to\infty}\frac{1}{T}\int_{0}^{T}\mathrm{d}t\,\mathrm{Tr}[\rho_{\alpha}e^{\mathrm{i}\mathcal{H}(\mathbf{k})t}\gamma_{d+1}e^{-\mathrm{i}\mathcal{H}(\mathbf{k})t}]\nonumber =-h_{d+1}(\mathbf{k})h_{\alpha}(\mathbf{k})/E^{2}(\mathbf{k}),\label{eb}
\end{align}
where $E(\mathbf{k})=\sqrt{\sum^{d+n}_{\alpha=1} h^2_\alpha}$ is the energy of the post-quenched Hamiltonian.
Besides, there are three aspects worth mentioning: (i) The determination of $\mathbf{k}^{(i-1)}$ obeys $h$-components in $\mathcal{B}^{(i-1)}_{\text{proj},s}$, but the calculation of $w_i$ is independent of the choice of $h$-components. (ii) When there are $\mathcal{B}_n$ in momentum space, the $(d-n)$D gapless boundary state must be emerged as long as the summation of topological charges in $\bar{\mathcal{B}}^{(n-1)}_{\text{proj},s}$ is non-zero. Here $\bar{\mathcal{B}}^{(i-1)}_{\text{proj},s}$ indicates the negative-value regions enclosed by the projective $s$th-order BISs $\mathcal{B}^{(i-1)}_{\text{proj},s}$, as it is shown in the main text.
(iii) We take $\mathcal{C}^{(i-1)}_{3,q}$ for $i<n$ (or $i<d-1$) and $\mathcal{C}^{(n-1)}_{2,q}$ [or $\mathcal{C}^{(d-2)}_{2,q}$ and $\mathcal{C}^{(d-1)}_{1,q}$] to facilitate the determination of $w_i$ for higher-order topological phases with $n<d$ (or $n=d$).

\begin{figure}[!ht]
\begin{center}
\includegraphics[width=\columnwidth]{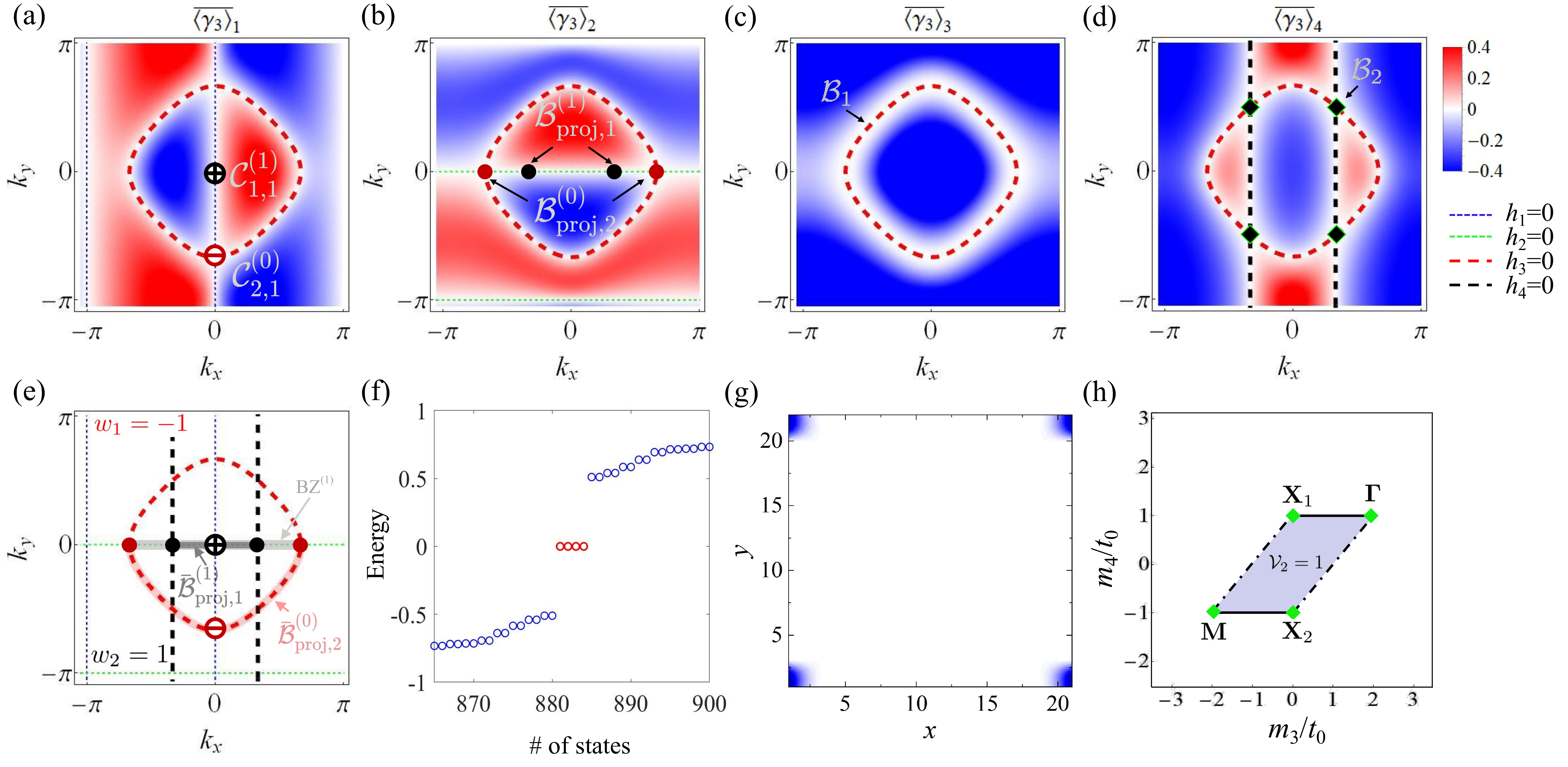}
\caption{Numerical results for the 2D 2nd-order TI. (a) $\overline{\langle\gamma_{3}(\mathbf{k}) \rangle}_{1}=0$
gives the 2nd-order (1st-order) topological charge $\mathcal{C}^{(0)}_{2,1}$ ($\mathcal{C}^{(1)}_{1,1}$) in $\text{BZ}^{(0)}$ ($\text{BZ}^{(1)}$).
(b) $\overline{\langle\gamma_{3}(\mathbf{k}) \rangle}_{2}=0$
gives the projective 2nd-order (1st-order) BISs $\mathcal{B}^{(0)}_{\text{proj},2}$ ($\mathcal{B}^{(1)}_{\text{proj},1}$) in $\text{BZ}^{(0)}$ ($\text{BZ}^{(1)}$).
(c) $\overline{\langle\gamma_{3}(\mathbf{k}) \rangle}_{3}=0$ presents a ring-shaped structure (red dashed ring), which gives the 1st-order BIS $\mathcal{B}_1$.
(d) $\overline{\langle\gamma_{3}(\mathbf{k}) \rangle}_{4}=0$ presents $h_3=0$ and $h_4=0$, which gives the 2nd-order BIS $\mathcal{B}_2$ (black diamonded points).
(e) A negative (positive) 2nd (1st)-order topological charge $\mathcal{C}^{(0)}_{2,1}$ ($\mathcal{C}^{(1)}_{1,1}$) lives on $\bar{\mathcal{B}}^{(0)}_{\text{proj},2}$ ($\bar{\mathcal{B}}^{(1)}_{\text{proj},1}$) and is within $\text{BZ}^{(0)}$ ($\text{BZ}^{(1)}$), giving $w_{1}=-1$ ($w_{2}=1$).
(f) $k_xk_y$-OBCs spectrum with four degenerate zero-energy states.
(g) Density distribution of zero-energy states.
(h) Phase diagram depends on the different $m_{3,4}$.
Here the other parameter is $t_{\text{so}}=t_0$.}
\label{Figs3}
\end{center}
\end{figure}
\begin{figure}[!ht]
\begin{center}
\includegraphics[width=\columnwidth]{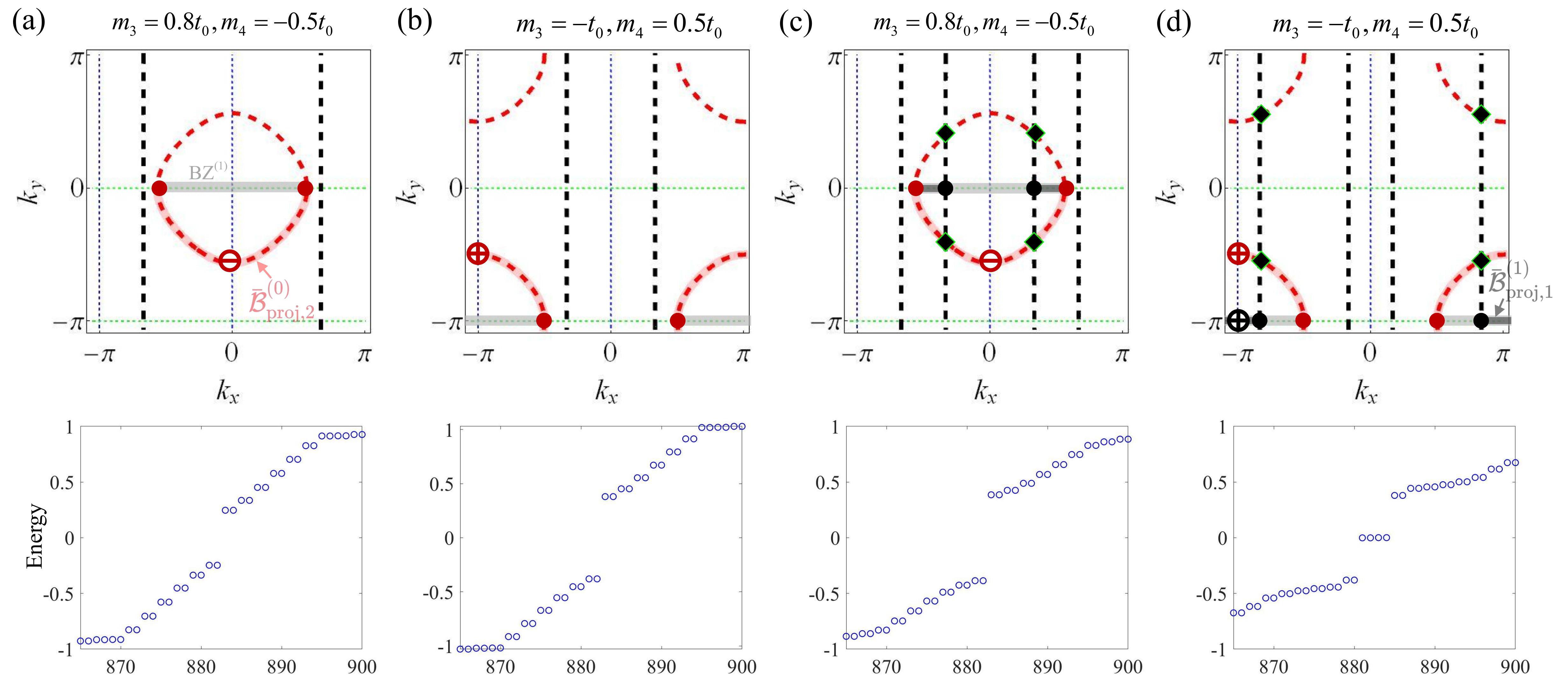}
\caption{Configurations of higher-order topological charges and the corresponding $k_xk_y$-OBC energy spectrum for the 2D 2nd-order TI, where (a) and (b) are for $h_4=m_4-t_0\cos{k_x}$, (c) and (d) are for $h_4=m_4-t_0\cos{(2k_x)}$. In (a) and (b), there is a 2nd-order topological charge in $\bar{\mathcal{B}}^{(0)}_{\text{proj},2}$, while no 2nd-order BISs $\mathcal{B}_2$ exists in BZ, which does not gives $\text{BZ}^{(1)}$.
In (c), although there are four 2nd-order BISs and gives $\text{BZ}^{(1)}$, there is no 1st-order topological charge on it. All of these give a trivial 2nd-order topological phase and no corner state.
In (d), there is a positive 2nd-order (1st-order) topological charge in $\bar{\mathcal{B}}^{(0)}_{\text{proj},2}$ ($\bar{\mathcal{B}}^{(2)}_{\text{proj},1}$), which holds a 2nd-order topological
phases and gives four zero-energy modes in each corner. Here the other parameter is $t_{\text{so}}=t_0$.}
\label{Figs4}
\end{center}
\end{figure}

\subsubsection{1. $2$D $2$nd-order topological insulators}

For the 2D 2nd-order TI described by the bulk Hamiltonian~(\ref{eqs10}), when the quench is performed by suddenly varying $(m_1,m_2,m_3,m_4)$ for each $h$-axis from $\delta m_{1,2,3,4}=30t_0$ to the post-quenched parameter $(0,0,0.5t_0,0.5t_0)$, we can obtain TASP of $\overline{\langle\gamma_{3}(\mathbf{k}) \rangle}_{1,2,3,4}$ by only detecting a single pseudospin component $\gamma_{3}$ in Fig.~\ref{Figs3}, where the vanishing TASP give $h_1=0,h_2=0,h_3=0$ and $h_4=0$. The 1st-order (red ring-shaped dashed line) and 2nd-order BISs (black square-shaped points) in original BZ are confirmed by $\overline{\langle\gamma_{3}(\mathbf{k}) \rangle}_{3,4}=0$ in Figs.~\ref{Figs3}(c) and \ref{Figs3}(d). Then we obtain the projective 2nd-order (1st-order) BISs $\mathcal{B}^{(0)}_{\text{proj},2}$ ($\mathcal{B}^{(1)}_{\text{proj},1}$)
in $\text{BZ}^{(0)}$ ($\text{BZ}^{(1)}$) by $\overline{\langle\gamma_{3}(\mathbf{k}) \rangle}_{2}=0$, which are presented by two red (black) points
in Fig.~\ref{Figs3}(b). Finally, we determine the corresponding 2nd-order (1st-order) topological charges $\mathcal{C}^{(0)}_{2,1}$ ($\mathcal{C}^{(1)}_{1,1}$) by TASP with $\overline{\langle\gamma_{3}(\mathbf{k}) \rangle}_{1}=0$, which are enclosed by $\mathcal{B}^{(0)}_{\text{proj},2}$ ($\mathcal{B}^{(1)}_{\text{proj},1}$) [see Fig.~\ref{Figs3}(a)].
From the vanishing TASP in Fig.~\ref{Figs3}(d), the existence of corner states is immediately identified, since we clearly observe the 2nd-order BISs $\mathcal{B}_2$ which present four point-shaped structure. A positive (negative) topological charge lives on $\bar{\mathcal{B}}^{(1)}_{\text{proj},1}$ ($\bar{\mathcal{B}}^{(0)}_{\text{proj},2}$) which is located on $\text{BZ}^{(1)}$ ($\text{BZ}^{(0)}$), giving $w_2=1$ ($w_1=-1$) in Fig.~\ref{Figs3}(e). Here $w_1$ and $w_2$ characterize the topology of $\mathcal{H}_{\mathbf{k}^{(0)}}=h_1\sigma_x+h_2\sigma_y+h_3\sigma_z$ and $\mathcal{H}_{\mathbf{k}^{(1)}}=h_1\sigma_y+h_4\sigma_z$, respectively. It should be noted that $\mathcal{H}_{\mathbf{k}^{(0)}}$ is now a block of $\mathbbm{Z}_2$-classified Hamiltonian $h_1\tau_z\otimes\sigma_x+h_2\tau_z\otimes\sigma_y+h_3\tau_z\otimes\sigma_z$ in $\tau_z=1$ eigenspace, and then its topology is $\mathbbm{Z}$-classified and can be identified by these topological charges. Correspondingly, this $\mathbbm{Z}_2$-classified Hamiltonian is characterized by $\text{sgn}|w_{1}|$ and can be recovered in $\mathcal{V}_2$. When taking $k_xk_y$-OBCs, the MDW-BIS duality is confirmed by the emergence of four degenerate zero-energy states which are indeed distributed on the four corners in Figs.~\ref{Figs3}(f) and \ref{Figs3}(g).
Besides, the phase diagram is shown in Fig.~\ref{Figs3}(h), where the existence of corner states is determined by the emergence of 2nd-order BISs $\mathcal{B}_2$, i.e., $|m_4|<t_0$ and $|m_3-m_4|<t_0$. We observe that bulk energy gap is closed at high-symmetry points $\mathbf{X}_1=(0,\pi)$, $\mathbf{X}_2=(\pi,0)$, $\boldsymbol{\Gamma}=(0,0)$, and $\mathbf{M}=(\pi,\pi)$, marked by the green squares. The solid and dot-dashed lines are determined by the higher-order topological charges living on the projective higher-order BISs and on the border of effective BZs respectively, in which only surface energy gap is closed. One can further see that the two types of HOTPTs can be captured by the location of topological charges, where the type-I HOTPT is that all higher-order topological charges pass through the projective higher-order BISs at the same time, thus the bulk gap is closed and then reopened. The type-II HOTPT is that only some higher-order topological charges cross over the projective higher-order BISs or the border of the effective BZs, then the bulk energy gap remains open but surface energy gap will close and then reopen.

Besides, we also take post-quenched parameters $m_3=-0.8t_0,m_4=-0.5t_0$ and $m_3=-t_0,m_4=0.5t_0$ to further illustrate our dynamical characterization.
For these two parameters, we observe that there is no projective 1st-order
BISs $\mathcal{B}^{(1)}_{\text{proj},1}$ and the corresponding 1st-order topological charge in $\text{BZ}^{(1)}$ (light-gray thick lines), as shown in Figs.~\ref{Figs4}(a) and \ref{Figs4}(b). The 2nd-order topology of system is trivial and no corner state exists, although the projective 2nd-order
BISs $\mathcal{B}^{(0)}_{\text{proj},2}$ and a negative (positive) 2nd-order topological charge exist in $\text{BZ}^{(0)}$ for (a) [(b)]. Especially, we change the mass term $h_4$ of Eq.~(\ref{eqs10}) into $h_4=m_4-t_0\cos (2k_x)$, and take the same parameters with (a) and (b). We observe that there are projective 1st-order
BISs $\mathcal{B}^{(1)}_{\text{proj},1}$ in $\text{BZ}^{(1)}$ (light-gray thick lines), as shown in Figs.~\ref{Figs4}(c) and \ref{Figs4}(d). However, only a positive 1st-order topological charge in the region $\bar{\mathcal{B}}^{(1)}_{\text{proj},1}$ (dark-gray thick lines) for Fig.~\ref{Figs4}(d), but there is no 1st-order topological charge in the region $\bar{\mathcal{B}}^{(1)}_{\text{proj},1}$ for Fig.~\ref{Figs4}(c). Therefore,
the 2nd-order TI is topological in Fig.~\ref{Figs4}(d) and there are corner states. While it is trivial in Fig.~\ref{Figs4}(c) and no corner state exists.
This results once again show that the existence of corner states is completely determined by the $n$-order BISs $\mathcal{B}_n$ in the original BZ, together with the non-zero invariant on a series of effective BZs.

\begin{figure}[!t]
\begin{center}
\includegraphics[width=\columnwidth]{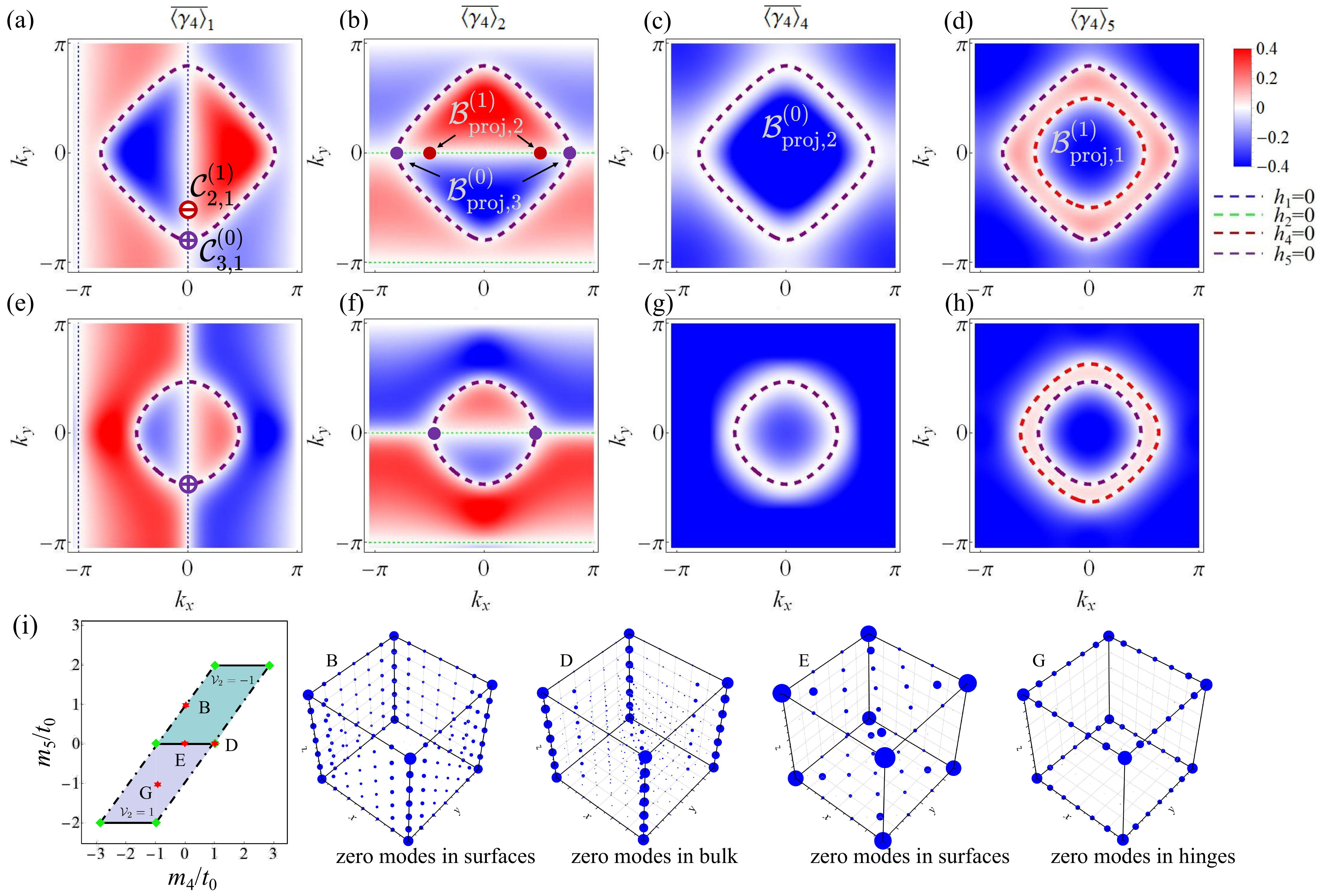}
\caption{Numerical results for the 3D 2nd-order TI.
(a) $\overline{\langle\gamma_{4}(\mathbf{k}) \rangle}_{1}=0$
gives the 3rd-order (2nd-order) topological charge $\mathcal{C}^{(0)}_{3,1}$ ($\mathcal{C}^{(1)}_{2,1}$) in $\text{BZ}^{(0)}$ ($\text{BZ}^{(1)}$).
(b) $\overline{\langle\gamma_{4}(\mathbf{k}) \rangle}_{2}=0$
gives the projective 3nd-order (2nd-order) BISs $\mathcal{B}^{(0)}_{\text{proj},3}$ ($\mathcal{B}^{(1)}_{\text{proj},2}$) in $\text{BZ}^{(0)}$ ($\text{BZ}^{(1)}$).
(c) $\overline{\langle\gamma_{4}(\mathbf{k}) \rangle}_{4}=0$ presents a ring-shaped structure (purple dashed ring), which gives the projective 2nd-order BIS $\mathcal{B}^{(0)}_{\text{proj},2}$ in $\text{BZ}^{(1)}$.
(d) $\overline{\langle\gamma_{5}(\mathbf{k}) \rangle}_{5}=0$ presents two ring-shaped structures, where one ring marked as red color gives the projective 1st-order BIS $\mathcal{B}^{(1)}_{\text{proj},1}$ in $\text{BZ}^{(1)}$.
(e)-(h) These TASP only give the 3rd-order topological charge $\mathcal{C}^{(0)}_{3,1}$ and the corresponding projective 2nd-order and 3rd-order BISs $\mathcal{B}^{(0)}_{\text{proj},2}$ and $\mathcal{B}^{(0)}_{\text{proj},3}$. There is no $\mathcal{C}^{(1)}_{2,1}$ and the 2nd-order topology of system is trivial.
(i) Phase diagram and the distribution of zero energy modes in real space, where the parameters are $(m_4,m_5)=(0,t_0)$ for $B$, $(t_0,0)$ for $D$, $(0,0)$ for $E$, and $(-t_0,-t_0)$ for $G$.
Here the other parameter is $t_{\text{so}}=t_0$.}
\label{Figs5}
\end{center}
\end{figure}

\subsubsection{2. $3$D $2$nd-order topological insulators}

For the 3D 2nd-order TI described by the bulk Hamiltonian~(\ref{eqs12}),
when the quench is performed by suddenly varying $(m_1,m_2,m_3,m_4,m_5)$ for each $h$-axis from $\delta m_{1,2,3,4,5}=30t_0$ to post-quenched parameters $(0,0,0,1.2t_0,t_0)$ with 2nd-order topology and $(0,0,0,2.1t_0,0.6t_0)$ without 2nd-order topology,
we obtain the TASP of $\overline{\langle\gamma_{4}(\mathbf{k}) \rangle}_{1,2,4,5}$ by only detecting a single pseudospin component $\gamma_{5}$ at $k_z=0$, as shown in Figs.~\ref{Figs5}(a)-\ref{Figs5}(d) and Figs.~\ref{Figs5}(e)-\ref{Figs5}(h) respectively.
From Figs.~\ref{Figs5}(a)-\ref{Figs5}(d), the vanishing TASP $\overline{\langle\gamma_{4}(\mathbf{k}) \rangle}_{4}=0$ presents a ring structure (purple dashed line) in Fig.~\ref{Figs5}(c), which gives the projective 2nd-order BIS $\mathcal{B}^{(0)}_{\text{proj},2}$ with $h_4=0$ and in $\text{BZ}^{(0)}$. The projective 1st-order BIS $\mathcal{B}^{(1)}_{\text{proj},1}$ in $\text{BZ}^{(1)}$ is the other ring-structure (red dashed line) with $h_5=0$, showing in Fig.~\ref{Figs5}(d).
In Fig.~\ref{Figs5}(b), the projective 3rd-order BISs $\mathcal{B}^{(0)}_{\text{proj},3}$
and projective 2nd-order BISs $\mathcal{B}^{(1)}_{\text{proj},2}$ are identified by $h_2=0$, $h_4=0$, and $h_5=0$.
In Fig.~\ref{Figs5}(a), a negative (positive) 2rd-order (3nd-order) topological charge $\mathcal{C}^{(1)}_{2,1}$ ( $\mathcal{C}^{(0)}_{3,1}$) is identified in $\text{BZ}^{(1)}$ ($\text{BZ}^{(0)}$) by $h_1=0$, $h_4=0$, and $h_5=0$.
Hence the topological charges $\mathcal{C}^{(0)}_{3,1}$ and $\mathcal{C}^{(1)}_{2,1}$ give $w_1=1$ and $w_2=-1$ respectively, determining the $2$nd-order topology $\mathcal{V}_2=-1$. Here $w_1$ and $w_2$ characterize the topology of $\mathcal{H}_{\mathbf{k}^{(0)}}=h_1\sigma_x\otimes \tau_0+h_2\sigma_y\otimes \tau_0+h_3\sigma_z\otimes \tau_x+h_4\sigma_z\otimes \tau_z$ and $\mathcal{H}_{\mathbf{k}^{(1)}}=h_1\sigma_x+h_2\sigma_y+h_5\sigma_z$, respectively. However, in Fig.~\ref{Figs5}(e)-\ref{Figs5}(h), we only observe a positive 3rd-order topological charge $\mathcal{C}^{(0)}_{3,1}=1$ while there no 2nd-order topological charge in $\text{BZ}^{(1)}$. Thus both $w_1=1$ and $w_2=0$ determine that the system has no hinge states and holds a trivial phase with $\mathcal{V}_2=0$.

\begin{figure}[!htb]
\begin{center}
\includegraphics[width=\columnwidth]{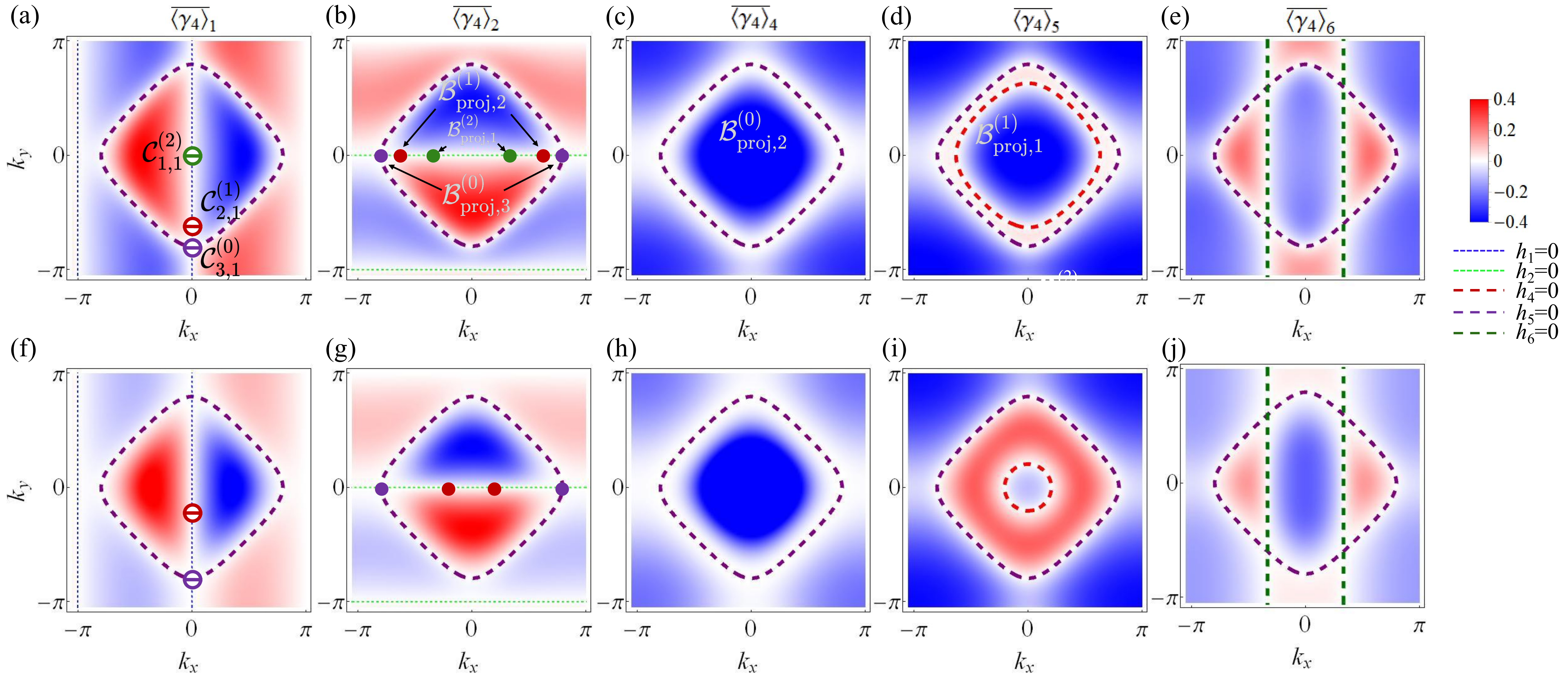}
\caption{Numerical results for the 3D 3nd-order TI.
(a) $\overline{\langle\gamma_{4}(\mathbf{k}) \rangle}_{1}=0$
gives the 3rd-order, 2nd-order, and 1st-order topological charge $\mathcal{C}^{(0)}_{3,1}$, $\mathcal{C}^{(1)}_{2,1}$, and $\mathcal{C}^{(2)}_{1,1}$ in $\text{BZ}^{(0)}$, $\text{BZ}^{(1)}$, and $\text{BZ}^{(2)}$ respectively.
(b) $\overline{\langle\gamma_{4}(\mathbf{k}) \rangle}_{2}=0$
gives the projective 3nd-order, 2nd-order, and 1st-order BISs $\mathcal{B}^{(0)}_{\text{proj},3}$, $\mathcal{B}^{(1)}_{\text{proj},2}$, and $\mathcal{B}^{(2)}_{\text{proj},1}$ in $\text{BZ}^{(0)}$, $\text{BZ}^{(1)}$, and $\text{BZ}^{(2)}$ respectively.
(c) $\overline{\langle\gamma_{4}(\mathbf{k}) \rangle}_{4}=0$ presents a ring-shaped structure (purple dashed ring), giving the projective 2nd-order BIS $\mathcal{B}^{(0)}_{\text{proj},2}$ in $\text{BZ}^{(0)}$.
(d) $\overline{\langle\gamma_{5}(\mathbf{k}) \rangle}_{5}=0$ presents two ring-shaped structures, where one ring marked as red color gives the projective 1st-order BIS $\mathcal{B}^{(1)}_{\text{proj},1}$ in $\text{BZ}^{(1)}$.
(e) $\overline{\langle\gamma_{5}(\mathbf{k}) \rangle}_{5}=0$ presents one ring-shaped and two line-shaped structures, where the ring gives the projective 2nd-order BIS $\mathcal{B}^{(0)}_{\text{proj},2}$ in $\text{BZ}^{(0)}$ and two line determine the projective 1st-order BIS $\mathcal{B}^{(2)}_{\text{proj},1}$ in $\text{BZ}^{(2)}$.
(f)-(j) The TASP only give the 3rd-order topological charge $\mathcal{C}^{(0)}_{3,1}$ and
2nd-order topological charge $\mathcal{C}^{(1)}_{2,1}$ in $\bar{\mathcal{B}}^{(0)}_{\text{proj},3}$ and $\bar{\mathcal{B}}^{(1)}_{\text{proj},2}$. There is no $\mathcal{C}^{(2)}_{1,1}$ and the 3rd-order TI is trivial.
Here the other parameter is $t_{\text{so}}=t_0$.}
\label{Figs6}
\end{center}
\end{figure}

Besides, we give the phase diagram in Fig.~\ref{Figs5}(i) again, where the existence of hinge states is determined by the emergence of 2nd-order BISs $\mathcal{B}_2$, i.e., $|m_5|<2t_0$ and $|m_4-m_5|<t_0$. For the different parameter points $B$, $D$, $E$, and $G$, the distribution of zero energy modes in real space further confirm two types of HOTPT points. For $B$, the topological charge $\mathcal{C}^{(1)}_{2,1}$ moves to the border of $\text{BZ}^{(1)}$, the $xz$ and $yz$ surfaces energy gap shall close, inducing $ \mathcal{V}_1=-1$. For $D$, all higher-order topological charges move to the projective higher-order BISs at the same time, the bulk energy gap closes and the system hosts $\mathcal{V}_2=0$. For $E$, the topological charge $\mathcal{C}^{(1)}_{2,1}$ moves to $\mathcal{B}^{(1)}_{\text{proj},2}$, the $xy$ surfaces energy gap shall close and drives $\mathcal{V}_1=-2$.
For $G$, only $\mathcal{C}^{(0)}_{3,1}$ moves to $\mathcal{B}^{(0)}_{\text{proj},3}$ but $\mathcal{C}^{(1)}_{2,1}$ still live in $\text{BZ}^{(1)}$, the system still host a $2$nd-order topological phase.

\subsubsection{3. $3$D $3$rd-order topological insulators}

\begin{figure}[!tb]
\begin{center}
\includegraphics[width=\columnwidth]{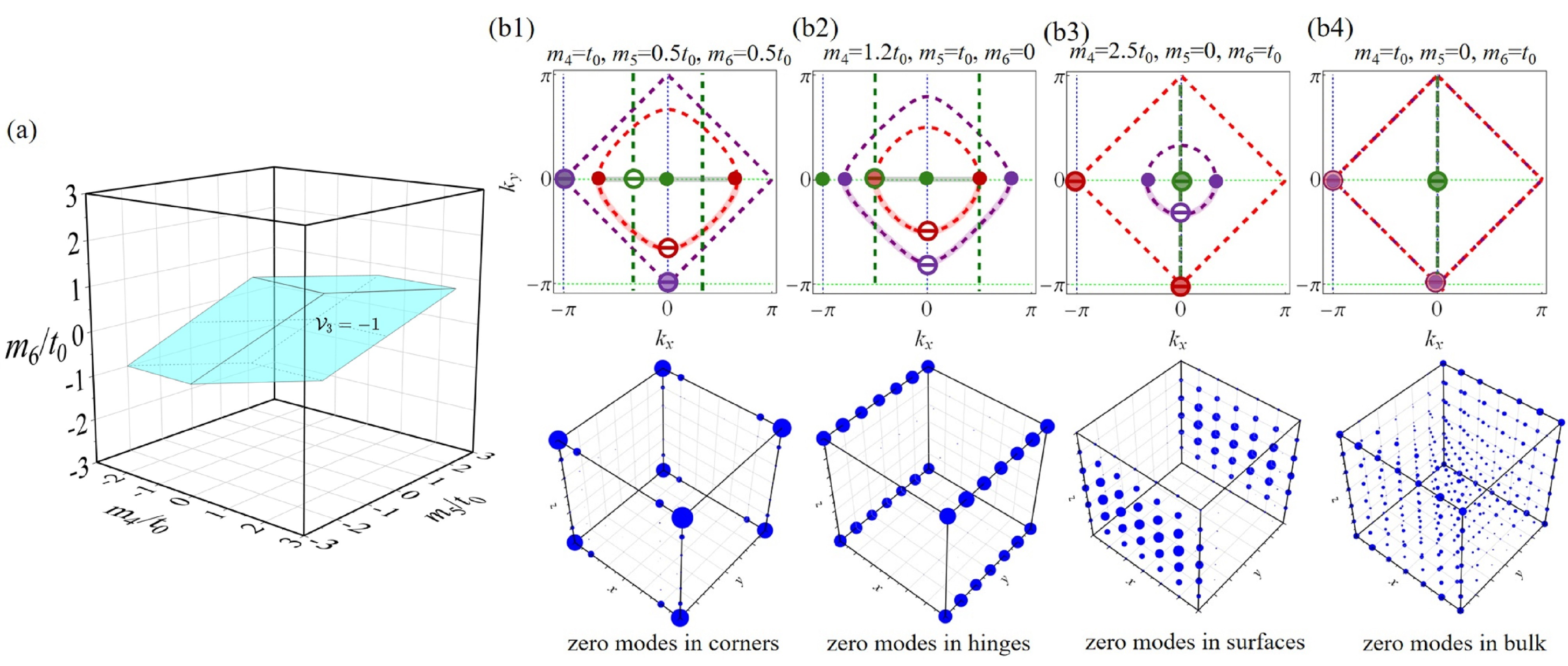}
\caption{Numerical results for the 3D 3nd-order TI.
(a) Phase diagram depends on the different $m_{4,5,6}$.
(b) Configurations of higher-order topological charges and projective higher-order BISs, where the corresponding distribution of zero energy modes in real space confirms the different HOTPT points. Here the other parameter is $t_{\text{so}}=t_0$.}
\label{Figs7}
\end{center}
\end{figure}

For the 3D 3nd-order TI described by the bulk Hamiltonian~(\ref{eqs14}),
when we quench the system by suddenly varying $(m_1,m_2,m_3,m_4,m_5,m_6)$ for each $h$-axis from $\delta m_{1,2,3,4,5,6}=30t_0$ to post-quenched parameters $(0,0,0,1.2t_0,0.6t_0,0.5t_0)$ with 3rd-order topology and $(0,0,0,1.2t_0,1.8t_0,0.5t_0)$ without 3rd-order topology, the TASP of $\overline{\langle\gamma_{4}(\mathbf{k}) \rangle}_{1,2,4,5,6}$ are obtained by only detecting a single pseudospin component $\gamma_{4}$ at $k_z=0$ in Figs.~\ref{Figs6}(a)-\ref{Figs6}(e) and Figs.~\ref{Figs6}(f)-\ref{Figs6}(j) respectively.
From Figs.~\ref{Figs6}(a)-\ref{Figs6}(e), we first identify
the projective 3rd-order BISs $\mathcal{B}^{(0)}_{\text{proj},3}$ (two purple points) in $\text{BZ}^{(0)}$, projective 2nd-order BISs $\mathcal{B}^{(1)}_{\text{proj},2}$ (two red points) in $\text{BZ}^{(1)}$ and projective
1st-order BISs $\mathcal{B}^{(2)}_{\text{proj},1}$ (two green points) in $\text{BZ}^{(2)}$, as shown in Fig.~\ref{Figs6}(b).
Further, we determine higher-order topological charges in Fig.~\ref{Figs6}(a), showing that there only is a negative 3rd-order topological charge $\mathcal{C}^{(0)}_{3,1}$ in $\bar{\mathcal{B}}^{(0)}_{\text{proj},3}$, 2nd-order topological charge $\mathcal{C}^{(1)}_{2,1}$ in $\bar{\mathcal{B}}^{(1)}_{\text{proj},2}$, and 1st-order topological charge $\mathcal{C}^{(2)}_{1,1}$ in $\bar{\mathcal{B}}^{(2)}_{\text{proj},1}$, respectively. Thus the bulk topology is given by $\mathcal{V}_3=-1$. Here $w_{1,2,3}$ characterize the topology of $\mathcal{H}_{\mathbf{k}^{(0)}}=h_1\sigma_y\otimes\tau_y+h_2\sigma_y\otimes\tau_z+h_3\sigma_z\otimes\tau_i+h_4\sigma_x\otimes\tau_i$ and $\mathcal{H}_{\mathbf{k}^{(1)}}=h_1\sigma_y+h_2\sigma_x+h_5\sigma_z$ and $\mathcal{H}_{\mathbf{k}^{(2)}}=h_1\sigma_y+h_6\sigma_z$. However, for Figs.~\ref{Figs6}(f)-\ref{Figs6}(j), we can identify that there are only projective 3rd-order BISs $\mathcal{B}^{(0)}_{\text{proj},3}$ in $\text{BZ}^{(0)}$ and projective 2nd-order BISs $\mathcal{B}^{(1)}_{\text{proj},2}$ in $\text{BZ}^{(1)}$, showing in Fig.~\ref{Figs6}(g). Although we also observe a negative 3rd-order (2nd-order) topological charge $\mathcal{C}^{(0)}_{3,1}$ ($\mathcal{C}^{(1)}_{2,1}$) in $\bar{\mathcal{B}}^{(0)}_{\text{proj},3}$ ($\bar{\mathcal{B}}^{(1)}_{\text{proj},2}$) in Fig.~\ref{Figs6}(f), there is no 1st-order topological charge $\mathcal{C}^{(2)}_{1,1}$ in $\bar{\mathcal{B}}^{(2)}_{\text{proj},1}$ and $\mathcal{V}_3=0$. Then there is no corner state in the system. Besides, we also give the phase diagram in Fig.~\ref{Figs7}(a), where the existence of corner states is determined by the emergence of 3nd-order BISs $\mathcal{B}_3$, i.e., $|m_6|<t_0$, $|m_4-m_5|<t_0$ and $|m_5-m_6|<t_0$. With closing bulk or surface or hinge energy gap, the different types of HOTPTs can occur.

All the previous dynamically numerical results are provided based on the definition of $\mathcal{B}^{(i-1)}_{\text{proj},s}\equiv\{\mathbf{k}^{(i)}|h_{d+i}=h_{d-i+1}=\cdots=h_{\beta+p}=0\}$ in main text. We take the corresponding $h$-components to define $\mathcal{B}^{(i-1)}_{\text{proj},s}$ and identify the higher-order topological charges $\mathcal{C}^{(i-1)}_{s,q}$. However, the determination of topological charge is independent of the choice of $h$-components and then $w_i$. Next we choose the different $h$-components to define topological charges $\mathcal{C}^{(2)}_{1,1}$ (i.e., $h_6=0$) and $\mathcal{B}^{(2)}_{\text{proj},1}$ (i.e., $h_1=0$) when deciding $w_3$, and discuss the two types of HOTPT points in Fig.~\ref{Figs7}(b). For the case of $(m_4,m_5,m_6)=(t_0,0.5t_0,t_0)$, $h_1=0$ gives the projective 1st-order BISs $\mathcal{B}^{(2)}_{\text{proj},1}$ (green points) in $\text{BZ}^{(2)}$ (gray thick line). A negative charge $\mathcal{C}^{(2)}_{1,1}$ (green color) in $\text{BZ}^{(2)}$ implies the system have corner states with $\mathcal{V}_3=-1$, as shown in Fig.~\ref{Figs7}(b1). When only the charge $\mathcal{C}^{(2)}_{1,1}$ moves to the border of $\text{BZ}^{(2)}$, the hinge energy gap shall close in Fig.~\ref{Figs7}(b2), inducing $\mathcal{V}_3=-1\rightarrow\mathcal{V}_2=-1$.
Further, when $\mathcal{C}^{(1)}_{2,1}$ (red color) and $\mathcal{C}^{(2)}_{1,1}$
simultaneously move to $\mathcal{B}^{(1)}_{\text{proj},2}$ (red points) and $\mathcal{B}^{(2)}_{\text{proj},1}$
while there is a negative $\mathcal{C}^{(0)}_{3,1}$ (purple color), the surface energy gap shall close in Fig.~\ref{Figs7}(b3), inducing $\mathcal{V}_3=-1\rightarrow\mathcal{V}_1=-2$. Finally, when all the higher-order topological charges simultaneously move to the projective higher-order BISs, the bulk energy gap shall close in Fig.~\ref{Figs7}(b4).

\bibliographystyle{apsrev4-1}

\begin{thebibliography}{103}
\bibitem [{not()}]{notes1}%
  \BibitemOpen
  \bibinfo {note} {For example of $\mathcal{H}_{\mathbf{k}^{(i-1)}}$
  perpendicular to $k_{d}\cdots k_{d-i+2}$-direction, it can be obtained by
  projecting $\mathcal{H}_\mathbf{k}$ with $n=i$ as
  $\mathcal{H}_{\mathbf{k}^{(i-1)}}=\mathbf{P}^{(i-2)^{\dagger}}\cdots
  \mathbf{P}^{(0)^{\dagger}}\mathcal{H}_\mathbf{k}\mathbf{P}^{(0)}\cdots
  \mathbf{P}^{(i-2)}$, where
  $\mathcal{H}_{\mathbf{k}^{(0)}}=\mathcal{H}_{\mathbf{k}}$. The projection
  operator is
  $\mathbf{P}^{(i-1)}=\left[1-\mathrm{i}\gamma^{(i-1)}_{d+i}\gamma^{(i-1)}_{d-i+1}\right]/2$
  in the eigenspace of
  $\mathrm{i}\gamma^{(i-1)}_{d+i}\gamma^{(i-1)}_{d-i+1}=-1$, where
  $\boldsymbol{\gamma}^{(i)}=\mathbf{P}^{(i-1)^{\dagger}}\boldsymbol{\gamma}^{(i-1)}\mathbf{P}^{(i-1)}$
  has the half size of $\boldsymbol{\gamma}^{(i-1)}$}\BibitemShut {NoStop}%
\end{thebibliography}

\end{document}